\RequirePackage{rotating}
\documentclass[12pt]{iopart}

\usepackage{graphicx}
\usepackage[normal]{subfigure}
\usepackage{caption}
\usepackage{latexsym}
\expandafter\let\csname equation*\endcsname\relax
\expandafter\let\csname endequation*\endcsname\relax
\usepackage{amsmath}
\usepackage{amssymb}
\usepackage{amsfonts}
\usepackage{color}
\usepackage{pgf, tikz}
\usepackage{dsfont}
\usepackage{caption}
\usepackage{amsthm}
\usepackage{mathtools}
\usepackage{array}
\usepackage{rotating,tabularx,booktabs,siunitx,caption}
\newcolumntype{C}{>{\centering\arraybackslash}X}

\usepackage[utf8]{inputenc}
\usepackage[english]{babel}
\usepackage[colorlinks]{hyperref}
\usepackage{etoolbox}

\makeatletter
\def\@mkboth#1#2{}
\newlength\appendixwidth
\preto\appendix{\addtocontents{toc}{\protect\patchl@section}}
\newcommand{\patchl@section}{%
	\settowidth{\appendixwidth}{\textbf{Appendix }}%
	\addtolength{\appendixwidth}{1.5em}%
	\patchcmd{\l@section}{1.5em}{\appendixwidth}{}{\ddt}%
}
\makeatother

\makeatletter
\newcommand{\mainmatter}{%
	\patchcmd{\@makefntext}{\fnsymbol}{\arabic}{}{}%
	\patchcmd{\@thefnmark}{\fnsymbol}{\arabic}{}{}%
	\def\@makefnmark{\textsuperscript{\arabic{footnote}}}%
}
\makeatother

\begin{document}

	\newcommand{\ket}[1]{\vert #1 \rangle}
\newcommand{\bra} [1] {\langle #1 \vert}
\newcommand{\braket}[2]{\langle #1 | #2 \rangle}
\newcommand{\proj}[1]{\ket{#1}\bra{#1}}
\newcommand{\mean}[1]{\langle #1 \rangle}
\newcommand{\opnorm}[1]{|\!|\!|#1|\!|\!|_2}
\newcommand{\mmmean}[1]{\langle\hspace{-3pt} \langle\hspace{-3pt} \langle #1 \rangle\hspace{-3pt} \rangle\hspace{-3pt} \rangle}
\newcommand{\mmean}[1]{\langle\hspace{-3pt} \langle #1 \rangle\hspace{-3pt} \rangle}
\newtheorem{theo}{Theorem}
\newtheorem{lem}{Lemma}
\newtheorem{defin}{Definition}
\newtheorem{corollary}{Corollary}
\newtheorem{conj}{Conjecture}
\newtheorem{prop}{Property}
\newtheorem{statement}{Statement}
\newtheorem{theobis}{Theorem}
\newtheorem{assum}{Assumption}
\newcommand{\kket}[1]{\vert\hspace{-2pt}\vert #1 \rangle \hspace{-3pt}\rangle}
\newcommand{\bbra} [1] {\langle \hspace{-3pt} \langle #1 \vert\hspace{-2pt}\vert}
\newcommand{\eq}[1]{\begin{equation}#1	\end{equation}}
\newcommand{\eqarray}[1]{\begin{eqnarray}#1	\end{eqnarray}}
\newcommand{\red}[1]{\textcolor{red}{#1}}
\newcommand{\x}{\hat{x}_1}
\newcommand{\xx}{\hat{x}_2}
\newcommand{\p}{\hat{p}_1}
\newcommand{\pp}{\hat{p}_2}
\newcommand{\theorem}[2]{\medskip \noindent\textbf{#1}:\\ \textit{#2} \medskip}

 \hypersetup{
 	linktocpage,
 	colorlinks,
 	linkcolor=blue,
 	citecolor=red,
 	urlcolor=blue
 }

\title[Continuous-variable entropic uncertainty relations]{Continuous-variable entropic uncertainty relations	
	\protect\footnote{Published in the special issue “{\it Shannon’s Information Theory 70 years on: Applications in classical and quantum physics}” in Journal of Physics A: Mathematical and Theoretical. Edited by Gerardo Adesso (Nottingham, UK), Nilanjana Datta (Cambridge, UK),  Michael Hall (Griffith, Australia), and Takahiro Sagawa (Tokyo, Japan). }
}

\author{Anaelle Hertz\footnote{Current address: Univ. Lille, CNRS, UMR 8523 - PhLAM - Physique des Lasers Atomes et
			Mol\'ecules, F-59000 Lille, France}}
\ead{ahertz@ulb.ac.be}
\address{Centre for Quantum Information and Communication, \'Ecole polytechnique de Bruxelles, Universit\'e libre de Bruxelles, 1050 Brussels, Belgium}

\author{Nicolas J. Cerf}
\ead{ncerf@ulb.ac.be}
\address{Centre for Quantum Information and Communication, \'Ecole polytechnique de Bruxelles, Universit\'e libre de Bruxelles, 1050 Brussels, Belgium}

\begin{abstract}
	Uncertainty relations are central to quantum physics. While they were originally formulated in terms of variances, they have later been successfully expressed with entropies following the advent of Shannon information theory. Here, we review recent results on entropic uncertainty relations involving continuous variables, such as position $x$ and momentum $p$. This includes  the generalization to arbitrary (not necessarily canonically-conjugate) variables as well as  entropic uncertainty relations that take $x$-$p$ correlations into account and admit all Gaussian pure states as minimum uncertainty states. We emphasize that these continuous-variable uncertainty relations can be conveniently reformulated in terms of entropy power, a central quantity in the information-theoretic description of random signals, which makes a bridge with variance-based uncertainty relations. In this review, we take the quantum optics viewpoint and consider uncertainties on the amplitude and phase quadratures of the electromagnetic field, which are isomorphic to $x$ and $p$, but the formalism applies to all such variables (and linear combinations thereof) regardless of their physical meaning. Then, in the second part of this paper, we move on to new results and introduce a tighter entropic uncertainty relation for two arbitrary vectors of intercommuting continuous variables that takes correlations into account. It is proven conditionally on reasonable assumptions. Finally, we present some conjectures for new entropic uncertainty relations involving more than two continuous variables.
\end{abstract}


\maketitle
\tableofcontents

\newpage

\mainmatter

	\makeatletter
	\long\def\@makefntext#1{%
		\parindent 1em\noindent \hb@xt@ 1.8em{\hss \normalfont\scriptsize\@thefnmark}.\enskip #1}
	\makeatother
\setcounter{footnote}{0}

\section{Introduction}

\label{section-introduction}

The uncertainty principle lies at the heart of quantum physics.  It exhibits one of the key discrepancies between a classical and a quantum system. Classically, it is in principle possible to specify the exact value of all measurable quantities in a given state of a system. In contrast, in quantum physics, whenever two observables do not commute, it is impossible to define a quantum state for which their values are simultaneously specified with infinite precision. 
First expressed by Heisenberg, in 1927, for position and momentum \cite{Heisenberg} it was formalized by Kennard \cite{Kennard} as
\begin{equation}
\sigma_x^2\sigma_p^2\geq \frac{\hbar^2}{4}
\end{equation}
where $\sigma_x^2$ and $\sigma_p^2$ denote the variance of
the position $x$ and momentum $p$, respectively, and $\hbar$ is the reduced Planck constant.
Shortly after, it was generalized to any pair of observables that do not commute \cite{Schrodinger,Robertson}. The uncertainty principle then states that their values cannot both be sharply defined beyond some precision depending on their commutator.

Aside from variances, another natural way of measuring the uncertainty of a random variable relies on entropy, the central quantity of Shannon information theory. In 1957, Hirschman stated the first entropic uncertainty relation \cite{Hirschman} but was only able to prove a weaker form of it. His conjecture was proven in 1975 independently by Białynicki-Birula and Mycielski \cite{Birula} and by Beckner \cite{Beckner}, making use of the work of Babenko \cite{Babenko}. It reads
\begin{equation}
h(x)+h(p)\geq\ln(\pi e \hbar)
\label{birulabis}
\end{equation} 
where $h(\cdot)$ is the Shannon differential entropy (see footnote$^{\ref{probdimension}}$ for a dimensionless version of this uncertainty relation). This result is interesting not only because it highlights the fact that Shannon information theory can help better understand fundamental concepts of quantum mechanics, but also because it opened the way to a new and fruitful formulation of uncertainty relations. Why such a success? First because Shannon entropy is arguably the most relevant measure of the degree of randomness (or uncertainty) of a random variable: it measures, in some asymptotic limit, the number of unbiased random bits needed to generate the variable or the number of unbiased random bits that can be extracted from it. In particular, building on Shannon's notion of entropy power, 
it can easily be seen that the entropic formulation of the uncertainty relation implies Heisenberg relation, so it is somehow stronger \cite{hertz2}.
In addition, unlike variance, the entropy is a relevant uncertainty measure even for quantities that are not associated with a numerical value or do not have a natural order. Moreover, entropic uncertainty relations can be generalized in a such way that (nonclassical) correlations with the environment are taken into account: typically, entanglement between a system and its environment can be exploited in order to reduce uncertainty.
If an observer has access to a quantum memory, the entropic formulation allows one to establish stronger uncertainty relations, which is particularly useful in quantum key distribution \cite{Renes,Berta}.
Uncertainty relations can then be used as a way to verify the security of a cryptographic protocol \cite{GrosshansCerf,Tomamichel1,Tomamichel2,Furrer}. 
They also find  applications in the context of  separability criteria, that is, criteria that enable one to distinguish between entangled and non-entangled states. For example, the positive-partial-transpose separability criterion for continuous variables \cite{duan,simon,walborn, huang2} is based on uncertainty relations: it builds on the fact that a state is necessarily entangled if its partial transpose is not physical, which itself is observed by the violation of an uncertainty relation. While refs. \cite{duan,simon} use variance-based uncertainty relations for this purpose, refs. \cite{walborn,huang2} exploit Shannon differential entropies (separability criteria can also be built with Rényi entropies \cite{rastegin}).
In general, a tighter uncertainty relation enables detecting more entangled states, hence finding better uncertainty relations leads to better separability criteria \cite{hertz}.



Somehow surprisingly, although entropic uncertainty relations were first developed with continuous variables, a large body of knowledge has accumulated over years on their discrete-variable counterpart. In a  seminal work, Deutsch proved in 1983 that $H(A)+H(B)$ has a nontrivial lower bound \cite{Deutsch}, where $H(\cdot)$ is the Shannon entropy and $A$ and $B$ are two incompatible discrete-spectrum observables. The lower bound was later improved by Kraus \cite{Kraus} and Maassen and Uffink \cite{MaassenUffink}, and much work followed on such uncertainty relations, with or without a quantum memory. We refer the reader to the recent review by Coles {\it et al.} \cite{colesReview}, where details on entropic uncertainty relations and their various applications can be found.

There is comparatively less available literature today on continuous-variable entropic uncertainty relations. Beyond ref. \cite{colesReview}, the older survey by Białynicki-Birula and Rudnicki \cite{Birula2} focuses on continuous variables but is missing the most recent results, while the recent review by Toscano {\it et al.} \cite{Toscano} is mainly concerned with coarse-grain measurements, which is a way to bridge the gap between discrete- and continuous-variable systems. With the present paper, we  provide an up-to-date  overview on continuous-variable entropic uncertainty relations that apply to any pair of canonically conjugate variables and linear combinations thereof. This review is meant to be balanced between the main results on this topic and some of our own recent contributions.

In Section \ref{introUR}, we first go over variance-based uncertainty relations as they serve as a reference for the entropic ones. In Section \ref{EURintro}, we review the properties of Shannon differential entropy as well as the notion of entropy power, and then move on to entropy-based uncertainty relations.  In particular, we define the entropic uncertainty relation due to Białynicki-Birula and Mycielski in Section \ref{EURforCCvariables}, and then introduce the entropy-power formulation which we deem appropriate to express continuous-variable uncertainty relations. Sections \ref{EURxpcorrelations} and \ref{EURcommutator} are  dedicated to more recent entropic uncertainty relations. In particular, the uncertainty relation of Section \ref{EURxpcorrelations} improves the Białynicki-Birula and Mycielski relation by taking $x$-$p$ correlations into account, and is then saturated by all pure Gaussian states. The entropic uncertainty relation of Section \ref{EURcommutator} is defined for any two vectors of intercommuting continuous variables, which are not necessarily related by a Fourier transform. In section \ref{otherEUR}, we briefly mention other possible variants of entropic uncertainty relations.
In the second part of this paper, we move on to new results and present in Section \ref{2EUR} a tight entropic uncertainty relation that holds for two vectors of intercommuting continuous variables. This relation is called tight because it is saturated by all pure Gaussian states. Finally, we propose in Section \ref{secConj} several conjectures in order to define an entropic uncertainty relation for more than two variables, and prove one of them. More than two observables have long been considered for variance-based uncertainty relations \cite{Robertson3,Weigert,dodonov2}, but, to our knowledge, no such result exists yet in terms of continuous entropies (except for a very recent conjecture by Kechrimparis and Weigert \cite{Weigert2}).

In \ref{appA}, we give a brief overview on Gaussian states and symplectic transformations, which should help readers who are less familiar with quantum optics to better understand this paper. \ref{appB} and \ref{appC} provide details on some calculations needed in Section \ref{2EUR}.



\section{Variance-based uncertainty relations}

\label{introUR} 

\subsection{Heisenberg-Kennard uncertainty relation}

In 1927, Heisenberg first  expressed an uncertainty relation between the position and  momentum of a particle. In a seminal paper \cite{Heisenberg}, he exhibited a thought experiment --- known as the Heisenberg's microscope --- for measuring the position of an electron. 
From this experiment, he concluded that there is a trade-off about how precisely the position $x$ and momentum $p$ can be both measured, which he expressed as 
$
\delta x \, \delta p \sim h ,
$
where $h$ is the Planck constant. 
Shortly after, Kennard \cite{Kennard} mathematically formalized the uncertainty relation and proved that
\begin{equation}
\sigma_x^2\sigma_p^2\geq \frac{\hbar^2}{4}
\label{heis}
\end{equation}
where $\sigma_x^2$ and $\sigma_p^2$ represent the variances  of the position and
momentum of a quantum particle and $\hbar=h/2\pi$ is the reduced Planck constant. 

Note that, as expressed by Kennard, the uncertainty relation is actually a property of Fourier transforms. 
While Heisenberg had made a statement about measurements, Kennard's formulation is really expressing an intrinsic property of the state. 
 Following Heisenberg's view, several papers have focused on finding an appropriate definition for measurement uncertainties (see \cite{BUSCH} for a review). In particular,  Ozawa \cite{ozawa} derived an inequality about error-disturbance and claimed that this is a rigorous version of Heisenberg’s formulation of the uncertainty principle. 
Nevertheless, this claim is still a matter of debate (for more details, see for example  \cite{Busch2,Busch3}).
Nowadays, most textbooks adopt the view of Kennard, as we do here, even though Eq. (\ref{heis}) is most often called the Heisenberg uncertainty relation.

\subsection{Schrödinger-Robertson uncertainty relation}
The uncertainty relation was originally formulated for position and momentum, but it is well known that it actually holds for any pair of canonically-conjugate variables, i.e., variables related to each other by a Fourier transform. For instance, the (amplitude and phase) quadrature components of a mode of the electromagnetic field are canonically-conjugate variables behaving just as position and momentum.\footnote{From now on, we consider these quadrature variables, also noted as $x$ and $p$, and do not make a distinction with their spatial counterparts. Thus, we take the quantum optics viewpoint on uncertainty relations and use the symplectic formalism in phase space, see \ref{appA}. We define, for example, uncertainty relations for $n$ modes, while they could address $n$ spatial degrees of freedom as well. Actually, the used formalism throughout the paper is quite general and applies to any canonically-conjugate variables (and linear combination thereof) regardless on their physical meaning.}  
Other canonical pairs can be defined, such as the charge and flux variables in a superconducting Josephson junction, verifying again Eq. (\ref{heis}).
In fact, in 1928, Robertson \cite{Robertson2}  extended the formulation of the uncertainty principle to any two arbitrary observables $\hat A$ and $\hat B$ as
\eq{ \sigma_A^2\sigma_B^2\geq \frac{1}{4}|\mean{\psi|[\hat A,\hat B]|\psi}|^2
	\label{rob}}
where $[\cdot,\cdot]$ stands for the commutator. 
Obviously, if $\hat A=\hat x$ and $\hat B= \hat p$, we recover Heisenberg uncertainty relation since $[\hat x,\hat p]=~i\hbar$. For simplicity, while being aware that uncertainty relations are expressed in terms of $\hbar$, we now fix $\hbar=1$.

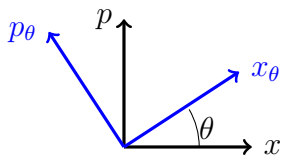
\begin{figure}[h!]
	\centering
	\begin{tikzpicture}
	\draw[very thick, ->] (0, 0) -- (1.7, 0) node[right]{$x$};
	\draw[very thick, ->] (0, 0) -- (0, 1.7) node[left]{$p$};
	\draw[blue, very thick, ->] (0, 0) -- (1.53, 1) node[right]{$x_\theta$};
	\draw[blue, very thick, ->] (0, 0) -- (-1, 1.53) node[left]{$p_\theta$};
	\draw (1, 0) arc (0: 30: 1cm);
	\node at (1.1,0.25) {$\theta$}; 
	\end{tikzpicture}
	\caption[Principle axes of the covariance matrix]{\label{axesprinc} Principal axes ($x_\theta,p_\theta$) of the covariance matrix $\gamma$, defined in such a way that $\sigma_{x_\theta\, p_\theta}=0$.}
\end{figure}

Relation (\ref{heis}) is invariant under  $(x,p)$-displacements in phase space since it only depends on central moments (esp. second-order moments of the deviations from the mean). Furthermore, it is saturated by all pure Gaussian states provided that they are squeezed in the $x$ or $p$ direction only.
More precisely, if we define the covariance matrix
\begin{equation}
\gamma = \begin{pmatrix}
\sigma_x^2 & \sigma_{xp} \\ 
\sigma_{xp} & \sigma_p^2 
\end{pmatrix} 
\end{equation}
where $\gamma_{ij}=\frac{1}{2}\mean{\{\hat r_i,\hat r_j\}}-\mean{\hat r_i}\mean{\hat r_j}$ and $\mathbf{r}=(\hat x,\hat p)$,
we see that Heisenberg relation is saturated by pure Gaussian states provided the principal axes of $\gamma$ are aligned with the $x$- and $p$-axes, namely  $\sigma_{xp}=0$.  The principal axes are defined as the $x_\theta$- and $p_\theta$-axes for which $\sigma_{x_\theta\, p_\theta}=0$, where 
\begin{equation}
\hat x_\theta = \cos \theta \, \hat x + \sin \theta \, \hat p,  \qquad\qquad
 \hat p_\theta = - \sin \theta \, \hat x + \cos \theta \, \hat p 
\end{equation}
are obtained by rotating $\hat x$ and $\hat p$ by an angle $\theta$ as shown in Figure \ref{axesprinc}.

The fact that Eq.~(\ref{heis}) is saturated only by certain pure Gaussian states is linked to the fact that this uncertainty relation is not invariant under rotations in phase space.
The problem of invariance was solved in 1930 by Schrödinger \cite{Schrodinger} and Robertson \cite{Robertson}, who added an anticommutator in relation~(\ref{rob}). The improved uncertainty relation for any two arbitrary observables then reads
\eq{\sigma_A^2\sigma_B^2\geq\frac{1}{4}\Big|\mean{\{A,B\}}-2\mean{A}\mean{B}\Big|^2+\frac{1}{4}\Big|\mean{[A,B]}\Big|^2
	\label{URwithanticomm}}
where $\mean{\cdot}$ is the shorthand notation for $\mean{\psi|\cdot|\psi}$.
In the special case of  position and momentum, $\hat A=\hat x$ and $\hat B=\hat p$,  the Robertson-Schrödinger uncertainty relation reads
\eq{\det\gamma\geq\frac{1}{4}.
	\label{robschr}}
This uncertainty relation is obviously invariant under symplectic transformations, i.e., squeezing and rotations (see \cite{weed} or \ref{appA} for more details on the symplectic formalism and phase-space representation), so it is saturated by all pure Gaussian states, regardless of the orientation of the principal axes of  $\gamma$.
Indeed, under a symplectic transformation $\mathcal{S}$, the new covariance matrix is given by $\gamma'=\mathcal{S}\gamma\mathcal{S}^T$. Since the determinant of a symplectic matrix is equal to 1, \eq{\det\gamma'=\det\mathcal{S}\det\gamma\det\mathcal{S}=\det\gamma}
which implies that Eq.~(\ref{robschr}) is invariant under symplectic transformations, hence under all Gaussian unitary transformations (since it is also invariant under displacements).

The generalization of the Robertson-Schrödinger uncertainty relation for the position and momentum variables of $n$ modes (or $n$ spatial degrees of freedom) is due to Simon et al. \cite{simon2}. It is formulated as an inequality on the covariance matrix $\gamma$
\eq{\gamma+\frac{i}{2}\Omega\geq0 \label{nmodeUR}}
where 
\eq{\Omega=\bigoplus\limits_{k=1}^n \omega,\qquad\qquad\omega=\begin{pmatrix}
		0&1\\-1&0	\end{pmatrix}.
}
 For one mode, Eq.~(\ref{nmodeUR}) reduces to the Robertson-Schrödinger uncertainty relation, but in general, we can  understand Eq.~(\ref{nmodeUR}) as $n$ inequalities that must be satisfied in order for the covariance matrix to represent a physical state. According to Williamson's theorem (see \ref{appA}), we can always diagonalize $\gamma$ in its symplectic form $\gamma^\oplus$ with the symplectic values $\nu_i$ on the diagonal (each $\nu_i$ appearing twice). Therefore, if $\gamma$ is the covariance matrix of a physical state, it  satisfies Eq.~(\ref{nmodeUR}) and so must $\gamma^\oplus$.
From this, we can show that  Eq. (\ref{nmodeUR}) is equivalent to (see \cite{these} for more details)
\eq{\nu_i\geq\frac{1}{2}\qquad\text{for }i=1,\cdots,n. \label{URsymplectic}}
Among others, an inequality that is easy to derive from Eq. (\ref{URsymplectic}) is  
\eq{\det\gamma=\det\gamma^\oplus=\prod_{i=1}^{n}\nu_i^2\geq\left(\frac{1}{4}\right)^n\label{RSanmodes}}
which is a straightforward $n$-mode generalization of the Robertson-Schrödinger uncertainty relation (\ref{robschr}).

\subsection{Uncertainty relation for more than two  observables}
Before concluding this section, let us mention that, in 1934, Robertson \cite{Robertson3} introduced a covariance-based uncertainty relation for $m$ observables which generalizes Eq. (\ref{URwithanticomm}). If we define the vector  $\mathbf{R}=(\hat R_1,\cdots,\hat R_m)$ of $m$  observables, then the uncertainty relation is expressed as 
\eq{\det\mathbf{\Gamma}\geq\det \mathbf {C} \label{robGen}}
where $\mathbf{\Gamma}$ is the covariance matrix of the measured observables and $\mathbf {C}$ the commutator matrix. Their elements are defined as 
\eq{\mathbf{\Gamma}_{ij}={1\over 2}\mean{\hat R_i\hat R_j+\hat R_j\hat R_i}-\mean{\hat R_i}\mean{\hat R_j},\qquad\qquad
	C_{ij}=-\frac{i}{2}\mean{[\hat R_i,\hat R_j]},\label{sigmaC} } 
respectively. For $m=2$, Eq. (\ref{robGen}) reduces to Eq. (\ref{URwithanticomm}). Surprisingly, when $m$ is odd, $\det \mathbf {C}=0$. Indeed, $\mathbf {C}$ is an antisymmetric matrix  ($C_{ij}=-C_{ji}$), so 
\eq{\mathbf{C}=-\mathbf{C}^T\quad\Leftrightarrow\quad\det \mathbf{C}=(-1)^m\det \mathbf{C}^T\quad\Leftrightarrow\quad\det \mathbf{C}=(-1)^m\det \mathbf{C} , }
which implies that this uncertainty relation is uninteresting for an odd number of observables.
For an even number of observables, $\det  \mathbf{C}$ is always non-negative \cite{Cayley}, so equation (\ref{robGen}) is interesting.
Note that unlike the situation with the Robertson-Schrödinger uncertainty relation, pure Gaussian states do not, in general, saturate Eq.~(\ref{robGen}). For more details on the minimum uncertainty states of this uncertainty relation, see \cite{trifonov}.

To circumvent the problem of this irrelevant bound for odd $m$, Kechrimparis and Weigert \cite{Weigert} proved in 2014 that for three pairwise canonical observables defined as $\hat p$, $\hat x$ and $\hat r=-\hat x-\hat p$ (which satisfy the commutation relations $[\hat p,\hat x]=[\hat x,\hat r]=[\hat r,\hat p]=-i$), the product of variances must satisfy the inequality
\begin{equation}
\sigma_x^2\sigma_p^2\sigma_r^2\geq\left(\frac{1}{\sqrt{3}}\right)^3.
\end{equation}
They later generalized this result to any vector $\mathbf{R}=(\hat R_1,\cdots,\hat R_m)$ of $m$ observables acting on one single mode as \cite{Weigert2} 
\begin{equation}
\sigma_1^2\sigma_2^2\cdots\sigma_m^2\geq\left(\frac{|\mathbf a\wedge \mathbf b|}{m}\right)^m
\label{eqStefan}
\end{equation}
where $\sigma_i^2=\mathbf{\Gamma}_{ii}$ are the variances of the $m$ observables,
$\mathbf a$ and $\mathbf b$ are defined through
\begin{equation}
	\mathbf R=\mathbf a \hat x+\mathbf b \hat p
	\label{defab}
\end{equation}
with $\hat x$ and $\hat p$ being the canonically conjugate quadratures of the mode, 
and the square norm of the wedge product $\mathbf a \wedge\mathbf b$ is computed as
\eq{|\mathbf a \wedge\mathbf b|^2=\sum_{i>j=1}^{m}(a_i b_j - a_j b_i)^2=|\mathbf a|^2|\mathbf b|^2-(\mathbf a \cdot\mathbf b)^2.\label{wedgeproduct}}
Shortly after, Dodonov also derived a general uncertainty relation involving any triple or quadruple of observables \cite{dodonov2}.

Note that Eq. (\ref{eqStefan}) takes a simple form in the special case where the $m$ one-modal observables are equidistributed quadratures over the unit circle, that is
\eq{\hat R_i=\cos\phi_i\, \hat x+\sin \phi_i \,\hat p\qquad\text{with}\quad\phi_i=\frac{2\pi(i-1)}{m},\qquad i=1,\dots,m.\label{defRpolygon}}
Indeed, the square norm of the wedge product $\mathbf a \wedge\mathbf b$ may be related to the matrix of commutators $\mathbf C$ as
\eq{|\mathbf a\wedge \mathbf b|^2=4\sum_{i>j=1}^{m}|C_{ij}|^2}
where the  $C_{ij}$ are  defined in Eq.~(\ref{sigmaC}). 
Then, for the observables $\hat R_i$ of Eq.~(\ref{defRpolygon}), it can be shown that
\eq{C_{ij}=\frac{1}{2}\sin\left(\frac{2\pi}{m}(j-i)\right)}
so that
\eq{|\mathbf a\wedge\mathbf b|^2=\sum_{i>j=1}^m\sin^2\left(\frac{2\pi}{m}(j-i)\right)=\frac{m^2}{4}   \label{eq-wedge-commutators}}
Plugging this into Eq. (\ref{eqStefan}) leads to the uncertainty relation \cite{Weigert2}
\eq{\sigma_1^2\sigma_2^2\cdots\sigma_m^2\geq\left(\frac{1}{2}\right)^m.\label{eqstefanpoly}}

\section{Entropy-based uncertainty relations}
\label{EURintro}

\subsection{Shannon differential entropy}

We start by reviewing the main properties of  Shannon differential (continuous-variable) entropy.
The differential entropy of a continuous (i.e., real-valued) variable $X$ with probability distribution $p(x)$ measures its uncertainty and
is defined as
\begin{equation}
h(X)\equiv h[p]=-\int_{-\infty}^{\infty}dx\,p(x)\ln p(x).
\label{diffEntropy}
\end{equation}
Here, the notation $h[p]$ implies that the entropy is a functional of the probability distribution $p(x)$, but it is often written $h(X)$ to stress that it refers to the random variable $X$. The definition (\ref{diffEntropy}) of the differential entropy is the natural continuous extension of the discrete entropy. More precisely, $h(X)$ is the limit of $H(X^\Delta)+\log\Delta$ when $\Delta\rightarrow0$, where $H(X^\Delta)$ is the discrete entropy of $X^\Delta$ defined as the discretized version of variable $X$ with discretization step $\Delta$. More details can be found in \cite{Cover, Birula2}.

For the probability distribution $p(x_1,\cdots,x_m)$ of $m$ continuous variables, we define the {\it joint} differential entropy of the vector $\mathbf{X}=(X_1,\cdots,X_m)$ as
\begin{equation}
h(\mathbf{X})=-\int dx_1\cdots dx_m\,p(x_1,\cdots,x_m)\ln p(x_1,\cdots,x_m).\label{diffEntropym}
\end{equation}
In addition, just like for discrete entropies, we may define the mutual information between two continuous variables $X_1$ and $X_2$ as
\eq{
	I(X_1{\rm :}X_2) =  h(X_1) + h(X_2) - h(X_1, X_2)    
.\label{mutualinfo}}
where $h(X_1, X_2)$ is the joint differential entropy and $h(X_1)$ and $h(X_2)$ are the differential entropies of the two marginals. The mutual information measures the shared entropy between $X_1$ and $X_2$ and is always non-negative.

Let us mention some useful properties of the differential entropy \cite{Cover}:
\begin{itemize}
	\item The differential entropy can be negative (unlike the discrete-variable entropy).
	\item The differential entropy is concave in $p(x)$.
\item The differential entropy is subadditive
\begin{equation}
h(\mathbf{X} ) \leq\sum_i h(X_i).\label{sub}
\end{equation}
\item  Under a translation, the value of the
differential entropy does not change
\begin{equation}
h(\mathbf{X} + \mathbf{c}) = h(\mathbf{X}),
\label{translation-h}
\end{equation}
where $\mathbf{c}$ is an arbitrary real vector.
\item Under a linear transformation, the differential entropy changes as
\begin{equation}
h(A\mathbf{X})=h(\mathbf{X})+\ln|\det A| \label{scalingRenyi},
\end{equation}
where $A$ is an invertible matrix that transforms the vector $\mathbf{X}$.
\end{itemize}

Note that the Shannon differential entropy actually belongs to the larger family of Rényi entropies. The Rényi entropy $h_\alpha(X)$ of parameter $\alpha$ is defined as
\begin{equation}
h_\alpha(X)=\frac{1}{1-\alpha}\log\left[\int_{-\infty}^{\infty}dx\,p^\alpha(x)\right].\label{renyientropy}
\end{equation}
and the limit of this expression when $\alpha\to 1$ converges to Shannon entropy, namely $\lim_{\alpha\to 1} h_\alpha(X) = h(X)$. Properties (\ref{translation-h}) and (\ref{scalingRenyi}) still hold for Rényi entropies, while it is not the case for concavity and subadditivity.

%

\subsection{Entropy power}
\label{entropypowersec}


Of particular interest is the entropy of a Gaussian distribution. Let  $\mathbf{X}=(X_1,\cdots,X_m)$ be a vector of $m$ Gaussian-distributed (possibly correlated) variables, \begin{equation}
p_G(\mathbf{x})=\frac{1}{\sqrt{(2\pi)^m \det \gamma}}e^{-\frac{1}{2}(\mathbf{x-\mean{x}})^T\gamma^{-1}(\mathbf{x-\mean{x}})}\label{GaussDistN}
\end{equation}
where $\mathbf{x}=(x_1\cdots,x_m)^T$ and $\gamma$ is the covariance matrix.
Its entropy is given by 
\eq{h(\mathbf{X)}={1\over2}\ln((2\pi e)^m\det\gamma). \label{entropygaussienne}}
For two Gaussian variables $X_1$ and $X_2$,  the  mutual information is given by
 \eq{I_G(X_1{\rm :}X_2)=\frac{1}{2}\ln\left(\frac{\sigma_1^2\sigma_2^2}{\det\gamma}\right)\label{gaussimutinfo}} where $\sigma_i^2$ is the variance of $X_i$ ($i=1,2$) and $\gamma$ is the covariance matrix of variables $X_1$ and $X_2$.

A key property of Gaussian distributions is that among all distributions $p(\mathbf x)$ with a same covariance matrix $\gamma$, the one having the maximum entropy is the Gaussian distribution $p_G(\mathbf x)$, that is
\begin{equation}
h[p]\leq h[p_G]={1\over2}\ln((2\pi e)^m\det\gamma).
\label{entropy-gausssian}
\end{equation}
Note that the equality is reached if and only if $p(\mathbf{x})$ is Gaussian.

From the subadditivity of the entropy applied to a multivariate Gaussian distribution, we get the {\it Hadamard inequality}
\begin{equation}
\det \gamma\leq\prod_i^m \sigma_i^2\label{hadam}  ,
\end{equation}
from which we can derive 
\eq{ \sigma_1^2\sigma_2^2\cdots\sigma_m^2\geq\det \mathbf C , \label{robweak}}
which is a weaker form of Robertson uncertainty relation (\ref{robGen}) for $m$ observables that ignores the correlations between them.


Now, exploiting (\ref{entropygaussienne}), we define the {\it entropy power} of a set of $m$ continuous random variables $\mathbf{X}=(X_1,\cdots,X_m)$ as
\begin{equation}
N_{\mathbf x}=\frac{1}{2\pi e}e^{\frac{2}{m}h(\mathbf{X})}  . 	\label{defentropypower}
\end{equation}
It is the variance\footnote{Although it is a variance, it is called ``power'' as it was introduced by Shannon in the context of the information-theoretic description of time-dependent signals.}  of a set of $m$ independent Gaussian variables that produce the same entropy as the set $\mathbf{X}$. The fact that the maximum entropy is given by a Gaussian distribution for a fixed covariance  matrix $\gamma$ translates, in terms of entropy powers, to
\eq{N_{\mathbf x}\leq(\det\gamma)^{1/m}.\label{maxGauss}}
%
For one variable, the entropy power is upper bounded simply by the variance, that is, $N_x\le \sigma_x^2$.
In the next section, we will show that the entropy power is a relevant quantity in order to express entropic uncertainty relations \cite{hertz2}. 


%

\subsection{Entropic uncertainty relation for canonically conjugate variables}
\label{EURforCCvariables}
The first formulation of an uncertainty  relation in terms of entropies is due to  Hirschman~\cite{Hirschman} in 1957. He conjectured an entropic uncertainty relation (EUR) for the position and momentum observables, which reads as follows:

\theorem{EUR for canonically-conjugate variables \cite{Birula,Beckner}}  {Any $n$-modal state $\rho$ satisfies the entropic uncertainty relation
	\begin{equation}
	h( \mathbf{x})+h(\mathbf{p})\geq n\ln (\pi e \hbar)
	\label{birulaEUR}
	\end{equation}
	where $\mathbf{x}=(\hat x_1,\cdots, \hat x_n)$ and $\mathbf{p}=(\hat p_1,\cdots, \hat p_n)$ are two vectors of pairwise canonically-conjugate quadratures \footnote{From now on, we make no precise distinction between the quadrature $\hat x$ ($\hat p$) and the random variable $X$ ($P$) that results from its measurement. The entropies of the random variables $X$ and $P$ will thus be noted $h(\hat x)$ and $h(\hat p)$, or simply $h(x)$ and $h(p)$.}  and $h(\cdot)$ is the differential entropy defined in Eq.~(\ref{diffEntropym}).  } 

Hirschman was only able to prove a weaker form of this conjecture (where $e$ is replaced by 2 in the lower bound) because of the known bound in the Hausdorff–Young inequality at the time.  The Hausdorff–Young inequality, which applies to Fourier transforms, is indeed at the heart of the proof of entropic uncertainty relations for canonically conjugate variables. A better bound was later found by Babenko \cite{Babenko} in 1961 and then by Beckner \cite{Beckner2} in 1975 (see also the work of Brascamp and Lieb \cite{brascamp}). This led to what is called the Babenko-Beckner inequality for Fourier transforms,
	\begin{equation}
 \left(\int d\mathbf{x} \, |\mathcal{F}f(\mathbf{x})|^p\right)^{1/p}\leq k(p,q) \left(\int d\mathbf{x} \, |f(\mathbf{x})|^q\right)^{1/q}
	\end{equation}
	where $\frac{1}{p}+\frac{1}{q}=1$, $k(p,q)=\left(\frac{2\pi}{p}\right)^{n/2p}\left(\frac{2\pi}{q}\right)^{-n/2q}$ and $\mathcal{F}f$ is the Fourier transform of function $f$. Using this last inequality,  Białynicki-Birula and Mycielski \cite{Birula} and independently Beckner \cite{Beckner} finally proved Eq. (\ref{birulaEUR}) in 1975.


Let us point out that Eq.~(\ref{birulaEUR}) may look weird at first sight as we take the logarithm of a quantity with dimension $\hbar$. This is a feature of the differential entropy itself since we have a similar issue in its definition, Eq.~(\ref{diffEntropy}), but the problem actually cancels out in Eq.~(\ref{birulaEUR})  
since we have dimension $\hbar$ on both sides of the inequality.\footnote{\label{probdimension}This problem was absent in the original expression of this uncertainty relation \cite{Birula} because the variable $k=p/\hbar$ was considered instead of $p$, giving $h(x)+h(k)~\ge ~\ln(\pi e)$ for $n=1$.} More rigorously, Eq.~(\ref{birulaEUR}) may be understood as the limit of a discretized version of the entropic uncertainty relation, with a discretization step tending to zero \cite{Birula2}.  
Being aware of this slight abuse of notation, we now prefer to keep $\hbar=1$ for simplicity.

As mentioned in \cite{Birula}, an interesting feature of inequality (\ref{birulaEUR}) is that it is stronger than -- hence it implies -- Heisenberg uncertainty relation, Eq. (\ref{heis}). This is easy to see if we formulate Eq. (\ref{birulaEUR}) in terms of entropy powers for one mode. Indeed, using Eq.~(\ref{defentropypower}), the  entropy powers of  $x$ and $p$ are defined as
\begin{equation}
N_x = {1\over 2\pi e} \, e^{2\, h(x)}, \qquad\qquad
N_p = {1\over 2\pi e} \, e^{2\, h(p)},
\label{def-of-ent-power-one-mode}
\end{equation}
so that the entropic uncertainty relation for one mode can be rewritten in the form of an entropy-power uncertainty relation \cite{hertz2}
\begin{equation}
N_x \, N_p \ge \frac{1}{4} ,
\label{EPUR}
\end{equation}
which closely resembles the Heisenberg relation (\ref{heis}) with $\hbar=1$.
Since $N_x \le \sigma_x^2$ and $N_p \le \sigma_p^2$, which reflects the fact that the Gaussian distribution maximizes the entropy for a fixed variance, we get the chain of inequalities
\begin{equation}
\sigma_x^2 \, \sigma_p^2 \ge N_x \, N_p \ge\frac{1}{4}
\end{equation}
so that Eq.~(\ref{EPUR}) implies the Heisenberg relation $\sigma_x^2\sigma_p^2\geq 1/4$. Note that since  $N_x=\sigma_x^2$  ($N_p=\sigma_p^2$)  if and only if $x$ ($p$) has a Gaussian distribution, the entropic uncertainty relation is strictly stronger than the Heisenberg relation for non-Gaussian states. As emphasized by Son \cite{Son}, the entropic uncertainty relation may indeed be viewed as an improved version of the Heisenberg relation where the lower bound is lifted up by exploiting an entropic measure of the non-Gaussianity of the state \cite{gauss4}, namely
\eq{\sigma_x^2\sigma_p^2\geq\frac{1}{4}\, e^{2\, D(x||x_G)+2\,D(p||p_G)}}
where $D(x||x_G)=h(x_G)-h(x)\geq0$ (and similarly for $p$) is the relative entropy between $x$ and $x_G$, namely the Gaussian-distributed variable with the same variance as $x$.

Just as the Heisenberg uncertainty relation, the entropy-power uncertainty relation (\ref{EPUR}) is only saturated for pure Gaussian states whose $\gamma$ has principal axes aligned with the $x$- and $p$-axes (i.e.,  $\sigma_{xp}=0$). It suggests that there is room for a tighter entropic uncertainty relation that is saturated for all pure Gaussian states, in analogy with the Robertson-Schrödinger uncertainty relation (\ref{robschr}). This is the topic of Section~\ref{EURxpcorrelations}.

As a final note, let us mention that one can also write an uncertainty relation for Rényi entropies as defined in Eq.~(\ref{renyientropy}). It reads as follows:

\theorem{Rényi EUR for canonically-conjugate variables \cite{birula3}}{ Any $n$-modal state $\rho$ satisfies the entropic uncertainty relation
\begin{equation}
h_\alpha(\mathbf x)+h_\beta(\mathbf p)\geq n\ln(\pi)+\frac{n\ln(\alpha)}{2\left(\alpha-1\right)}+\frac{n\ln(\beta)}{2\left(\beta-1\right)}
\label{EURrenyi}
\end{equation}
where $\mathbf{x}=(\hat x_1,\cdots, \hat x_n)$ and $\mathbf{p}=(\hat p_1,\cdots, \hat p_n)$ are two vectors of pairwise canonically-conjugate quadratures and $h_\alpha(\cdot)$ is the Rényi entropy defined in Eq.~(\ref{renyientropy}), with parameters $\alpha$ and $\beta$ satisfying
\begin{equation}
\frac{1}{\alpha}+\frac{1}{\beta}=2.
\end{equation}}

In \cite{Jizba}, the entropy-power formulation associated with Rényi entropies was used to show that some Gaussian states saturate these entropic uncertainty relations for all parameter $\alpha$ and $\beta$ such that $\frac{1}{\alpha}+\frac{1}{\beta}=2$ . However, for some parameters,  it is possible to find non-Gaussian states that saturate them too. For more information about entropic uncertainty relations with Rényi entropies, see also refs. \cite{birula2007,rastegin2015,Rudnicki}.

\subsection{Tight entropic uncertainty relation for canonically conjugate variables} 

\label{EURxpcorrelations} 

The  entropic uncertainty relation, Eq.~(\ref{birulaEUR}), is not invariant under all symplectic transformations and is not saturated by all pure Gaussian states. However, a tighter entropic uncertainty relation can be written, which, by taking correlations into account, becomes saturated for \textit{all} Gaussian pure states. It is expressed as follows.

\theorem {Tight EUR for canonically-conjugate variables \cite{hertz2}} {Any $n$-modal state $\rho$  satisfies the entropic uncertainty relation \footnote{The proof is conditional on two reasonable assumptions, see below and \cite{hertz2}.}
\begin{equation}
h(\mathbf x)+h(\mathbf p)   -  {1\over 2} \ln \left(\frac{\det\gamma_{\mathbf{x}}\det\gamma_{\mathbf{p}}}{  \det \gamma} \right) \ge n\,\ln(\pi e )
\label{nmodes}
\end{equation}
where $\mathbf{x}=(\hat x_1,\cdots, \hat x_n)$ and $\mathbf{p}=(\hat p_1,\cdots, \hat p_n)$ are two vectors of pairwise canonically-conjugate quadratures and $h(\cdot)$ is the differential entropy defined in Eq.~(\ref{diffEntropym}). The covariance matrix $\gamma$ is defined as
$\gamma_{ij}=\mathrm{Tr}[\hat \rho\, \{r_i,r_j\}]/2-\mathrm{Tr}[\hat\rho \, r_i]\mathrm{Tr}[\hat\rho \, r_j]$ with $\mathbf{r}~=~(\hat x_1,\cdots,\hat x_n,\hat p_1,\cdots,\hat p_n)$, and $\gamma_{\mathbf{x}}$ ($\gamma_{\mathbf{p}}$) denotes the reduced covariance matrix of the $\mathbf{x}$ ($\mathbf{p}$) quadratures. Eq.~(\ref{nmodes}) is saturated if and only if $\rho$ is Gaussian and pure.   }



In the context of entropic uncertainty relations, it would be natural to take correlations into account via the joint entropy $h(x,p)$ of the two canonically-conjugate quadratures $x$ and $p$ (considering first the case of a single mode, $n=1$). The problem, however, is that $h(x,p)$  is not defined for states with a negative Wigner function (more details can be found in \cite{hertz2,these}). To overcome this problem, correlations can be accounted for by exploiting the covariance matrix $\gamma$. Indeed, the mutual information $I(x{\rm :}p)$ between two Gaussian variables ($x$ and $p$ for a one-modal state) can be expressed in terms of  the covariance matrix, see Eq.~(\ref{gaussimutinfo}). Then, starting from the joint entropy $h(x,p)=h(x)+h(p)-I(x{\rm :}p)$  and substituting $I(x{\rm :}p)$ by its Gaussian form, Eq. (\ref{gaussimutinfo}), we get a quantity that is defined for all states regardless of whether the Wigner function is positive or not. This yields a tight entropic uncertainty relation \cite{hertz2}
\begin{equation}
h(x)+h(p) -  {1\over 2} \ln \left(  \frac{\sigma_x^2\sigma_p^2}{\det\gamma}\right)  \ge \ln(\pi e )
\label{CONJECTURE}
\end{equation}
whose generalization to $n$ modes corresponds to Eq. (\ref{nmodes}). Thus, the lower bound of the entropic uncertainty relation (\ref{birulaEUR}) can be lifted up by a non-negative term that exploits the covariance matrix $\gamma$. 


The entropic uncertainty relation (\ref{nmodes}) applies to any state, Gaussian or not. For Gaussian states, it is easy to prove. Indeed, using Eq.~(\ref{entropygaussienne}) and  $\det\gamma\geq1/4^n$ [which is simply the  $n$-modal version of Robertson-Schrödinger uncertainty relation, Eq. (\ref{RSanmodes})], we have
\eqarray{h(\mathbf x)+h(\mathbf p) -  {1\over 2} \ln \left(\frac{\det\gamma_{\mathbf{x}}\det\gamma_{\mathbf{p}}}{  \det \gamma} \right) 
	&=&n\ln(\pi e) +\frac{1}{2}\ln(4^n\det\gamma)\nonumber\\
	&\geq&n\ln(\pi e)
}
This inequality is saturated if and only if the state is pure since $\det\gamma=1/4^n$ for pure Gaussian states only. Thus, Eq.~(\ref{nmodes}) is a tight uncertainty relation in the sense that it is
 saturated for all pure Gaussian states, regardless of the orientation of the principal axes. Nevertheless, Eq.~(\ref{nmodes}) is not invariant under rotations.

For non-Gaussian states, the proof of Eq.~(\ref{nmodes}) is more involved and only partial. We do not give the full details here as they can be found in  \cite{hertz2} (or in Section \ref{2EUR}, where we use the same technique of proof). In a nutshell, the proof relies on a variational method similar to the procedure used in Ref.~\cite{jackiw, Hall}. One defines the uncertainty functional 
\begin{equation}
F(\hat\rho) = 	h(\mathbf x)+h(\mathbf p)   -  {1\over 2} \ln \left(\frac{\det\gamma_{\mathbf{x}}\det\gamma_{\mathbf{p}}}{  \det \gamma} \right).
\label{uncertainty-functional}
\end{equation}
and shows that any $n$-modal squeezed vacuum state is a local extremum of $F(\hat\rho)$. Since $F(\hat\rho)$ is invariant under $(x,p)$-displacements, it follows that all Gaussian pure states are extrema too. To complete the proof of Eq.~(\ref{nmodes}), one must take the two following statements for granted:
\begin{enumerate}
\item \label{assum1}Pure Gaussian states are global minimizers of the uncertainty functional $F(\hat \rho)$.
\item  \label{assum2}The uncertainty functional $F(\hat \rho)$ is concave, so relation (\ref{nmodes}) is  valid. 
\end{enumerate} 
Remark that (i) and (ii) both prevail for the uncertainty functional $h(\mathbf x)+h(\mathbf p)$ appearing in the entropic uncertainty relation (\ref{birulaEUR}).

For one mode,  the entropy-power formulation of Eq.~(\ref{CONJECTURE}) reads
\begin{equation}
{N_x \, N_p\over \sigma_x^2 \, \sigma_p^2} ~  \det \gamma\ge \frac{1}{4},
\label{tight-entropy-power-UR}
\end{equation}
where $N_x$ and $N_p$ are the entropy powers defined in Eq. (\ref{def-of-ent-power-one-mode}).
This highlights the fact that the Robertson-Schrödinger relation (\ref{robschr}) can be deduced from the tight entropy-power uncertainty relation (\ref{tight-entropy-power-UR}). Indeed, since $N_x \le \sigma_x^2$ and $N_p \le \sigma_p^2$, we have the chain of inequalities
\begin{equation}
\det \gamma\ge  {N_x \, N_p\over \sigma_x^2 \, \sigma_p^2} ~  \det \gamma \ge  \frac{1}{4}
\end{equation}
and, once again,  both inequalities coincide only for Gaussian $x$- and $p$-distributions. 


For $n$ modes, the entropy-power formulation of Eq.~(\ref{nmodes}) becomes
\begin{equation}
\frac{\left( N_{\mathbf{x}} N_{\mathbf{p}} \right)^n}{\det \gamma_{\mathbf{x}}\, \det \gamma_{\mathbf{p}}} \, \det \gamma \geq  \left(\frac{1}{4}  \right)^{n} 
\label{n-mode-conjecture-bis}
\end{equation}
where
\begin{equation}
N_{\mathbf{x}}=\frac{1}{2\pi e}e^{{2\over n}h(\mathbf x)}\qquad\qquad 
N_{\mathbf{p}}=\frac{1}{2\pi e}e^{{2\over n}h(\mathbf p)}.
\end{equation}
Here too, we can use the fact that the maximum entropy for a fixed covariance matrix is given by the Gaussian distribution, which implies that
\begin{equation}
\det \gamma\geq\frac{\left( N_{\mathbf{x}} N_{\mathbf{p}} \right)^n}{\det \gamma_{\mathbf{x}}\, \det \gamma_{\mathbf{p}}} \, \det \gamma \geq  \left(\frac{1}{4}  \right)^{n} 
\end{equation}
that is, the  $n$-mode tight entropy-power uncertainty relation (\ref{n-mode-conjecture-bis}) implies the $n$-mode variance-based Robertson-Schrödinger uncertainty relation, Eq. (\ref{RSanmodes}).


\subsection{Entropic uncertainty relation for arbitrary quadratures}
\label{EURcommutator} 

Traditionally, continuous-variable entropic uncertainty relations have been formulated for the position and momentum quadratures or, more precisely, for continuous variables that are related by a Fourier transform. However, as for variance-based ones, entropic uncertainty relations can be extended to any pair of variables. In 2009, Guanlei \emph{ et al.} \cite{Guanlei} first formulated an entropic uncertainty relation for two rotated quadratures:

\theorem{EUR for two rotated quadratures \cite{Guanlei}}{ Any one-mode state $\rho$ satisfies the entropic uncertainty relation
\begin{equation}
h(x_\theta)+h(x_\phi)\geq \ln(\pi e | \sin(\theta-\phi)|).
\label{eq-Guanlei}
\end{equation}
where 
$\hat{x}_\theta=\hat{x} \cos \theta + \hat{p} \sin \theta$ and $\hat{x}_\phi= \hat{x} \cos \phi + \hat{p} \sin \phi$ are two rotated quadratures, and $h(\cdot)$ is the Shannon differential entropy.}

In 2011, Huang \cite{huang} obtained a more general entropic uncertainty relation that holds for any pair of observables, that is, two variables that are not necessarily canonically conjugate (or that are not related by a Fourier transform): 

\theorem{EUR for two arbitrary quadratures \cite{huang}}{ Any $n$-modal state $\rho$ satisfies the entropic uncertainty relation
\begin{equation}
h(\hat{A})+h(\hat{B})\geq\ln(\pi e |[\hat{A},\hat{B}]|)
\label{huang}
\end{equation}
where $h(\cdot)$ is the Shannon differential entropy, $\hat A$ and $\hat B$ are two observables defined as 
\begin{equation}
\hat{A}=\sum_{i=1}^n(a_i \, \hat{x}_i+a_i' \, \hat{p}_i),\quad
\hat{B}=\sum_{i=1}^n(b_i \, \hat{x}_i+b_i' \, \hat{p}_i),
\label{ABdef}
\end{equation} 
and $[\hat{A},\hat{B}]$ (which is a scalar) is the commutator between them.}

 Obviously, if $\hat{A}=\hat{x}$ and $\hat{B}=\hat{p}$, this inequality reduces to the entropic uncertainty relation of Białynicki-Birula and Mycielski, Eq.~(\ref{birulabis}),
while it reduces to Eq.~(\ref{eq-Guanlei}) if $n=1$. 

More recently, an entropic uncertainty relation that holds for any two vectors of not-necessarily canonically conjugated variables was derived in \cite{hertz3}. The bound on entropies  is then expressed in terms of the determinant of a $n\times n$ matrix formed with the commutators between the $n$ measured variables:

\theorem{EUR for two arbitrary vectors of intercommuting quadratures \cite{hertz3}}{ 
	Let $\mathbf y = (\hat y_1,\cdots \hat y_n)^T$ be a vector of commuting quadratures and $\mathbf z =~(\hat z_1,\cdots \hat z_n)^T$ be another vector of commuting quadratures. Let each of the components of $\mathbf y$ and $\mathbf z$ be written as a linear combination of the $(\hat x, \hat p)$ quadratures of an $n$-modal system, namely
	\begin{eqnarray}
	\hat y_i &=& \sum_{k=1}^{n} a_{i,k} \, \hat x_k  + \sum_{k=1}^{n} a'_{i,k} \, \hat p_k  \qquad (i=1,\cdots n) \nonumber \\
	\hat z_j &=& \sum_{k=1}^{n} b_{j,k} \, \hat x_k  + \sum_{k=1}^{n} b'_{j,k} \, \hat p_k   \qquad (j=1,\cdots n).
	\label{quadyz}
	\end{eqnarray}
	Then, any $n$-modal state $\rho$ satisfies the entropic uncertainty relation
	\begin{equation}
	h(\mathbf y)+h(\mathbf z)\geq\ln\left((\pi e)^n |\det \mathbf{K}|\right)
	\label{EURFRFTcom}
	\end{equation}
	where  $h(\cdot)$ stands for the Shannon differential entropy of the probability distribution of the vectors of jointly measured quadratures $\hat y_i$'s or $\hat z_j$'s, 
	and  $\mathbf{K}_{ij}=[\hat y_i,\hat z_j]$ denotes the $n\times n$ matrix of commutators (which are scalars).
}

The proof of Eq.~(\ref{EURFRFTcom}) exploits the fact that the probability distributions of vectors $\mathbf y$ and $\mathbf z$ are related by a fractional Fourier transform (instead of a simple Fourier transform). Then, the  entropic  uncertainty relation of Białynicki-Birula and Mycielski, Eq.~(\ref{birulaEUR}), simply corresponds to the special case of Eq.~(\ref{EURFRFTcom}) for a Fourier transform, that is, when measuring either all $x$ quadratures or all $p$ quadratures on $n$ modes. Also, Eqs.~(\ref{eq-Guanlei}) and (\ref{huang}) are special cases of Eq.~(\ref{EURFRFTcom}) for a one-by-one matrix $\mathbf{K}$.
Finally, let us also mention that Eq.~(\ref{EURFRFTcom}) still holds if we jointly measure $n$ quadratures on a larger $N$-dimensional system i.e. when the sum over $k$  in equation (\ref{quadyz}) goes to $N$  (with $N>n$) \cite{hertz3}.

%

By exploiting the entropy-power formulation of Eq.~(\ref{EURFRFTcom}), it is  possible to derive an $n$-dimensional extension of the usual Robertson uncertainty relation in position and momentum spaces where, instead of expressing the complementarity between observables $\hat A$ and $\hat B$ (which are linear combinations of quadratures, so the commutator $ [\hat A,\hat B]$ is a scalar), 
one expresses the complementarity between two vectors of intercommuting observables. Defining the entropy powers of $\mathbf y$ and $\mathbf z$ as
\begin{equation}
N_{\mathbf{y}}=\frac{1}{2\pi e}e^{\frac{2}{n}h(\mathbf y)},\qquad\quad
N_{\mathbf{z}}=\frac{1}{2\pi e}e^{\frac{2}{n}h(\mathbf z)},
\label{Nyz}
\end{equation}
we can rewrite Eq.~(\ref{EURFRFTcom}) as an entropy-power uncertainty relation for two arbitrary vectors of intercommuting quadratures $\mathbf y$ and $\mathbf z$, namely
\begin{equation}
N_{\mathbf{y}}N_{\mathbf{z}}\geq \frac{ |\det \mathbf{K}|^{2/n} }{4} .
\label{entropy-power-ur}
\end{equation} 
Again, we may use the fact that the maximum entropy for a fixed covariance matrix is reached by the Gaussian distribution and write $N_{\mathbf{y}}\leq~(\det \mathbf{\Gamma}_{\mathbf{y}})^{1/n}$ and $N_{\mathbf{z}}\leq~(\det \mathbf{\Gamma}_{\mathbf{z}})^{1/n}$, where $(\mathbf{\Gamma}_{\mathbf{y}})_{ij}=\langle\{\hat y_i,\hat y_j\}\rangle/2-\langle\hat y_i\rangle\langle\hat y_j\rangle$ and $(\mathbf{\Gamma}_{\mathbf{z}})_{ij} =~\langle\{\hat z_i,\hat z_j\}\rangle/2-\langle\hat z_i\rangle\langle\hat z_j\rangle$ are the (reduced) covariance matrices of the $\hat y_i$ and $\hat z_i$ quadratures. 
Combining these inequalities with Eq.~(\ref{entropy-power-ur}), we obtain the $n$-modal variance-based uncertainty relation (VUR): 

\theorem{VUR for two arbitrary vectors of intercommuting quadratures \cite{hertz3}}{
	Let $\mathbf y = (\hat y_1,\cdots \hat y_n)^T$ be a vector of commuting quadratures, $\mathbf z =~(\hat z_1,\cdots \hat z_n)^T$ be another vector of commuting quadratures, and let each of the components of these vectors be written as a linear combination of the $(\hat x, \hat p)$ quadratures of an $N$-modal system ($N \ge n$).  Then,
any $N$-modal state $\rho$ verifies the variance-based uncertainty relation	\begin{equation}
	 \det \mathbf{\Gamma}_{\mathbf{y}} \,  \det \mathbf{\Gamma}_{\mathbf{z}}   \geq\frac{|\det \mathbf{K}|^2}{4^n}
	\label{URcovariance}
	\end{equation}
where $ \mathbf{\Gamma}_{\mathbf{y}}$  ($\mathbf{\Gamma}_{\mathbf{z}}$) is the covariance matrix of the jointly measured quadratures $\hat y_i$'s ($\hat z_j$'s), and
$\mathbf{K}_{ij}=[\hat y_i,\hat{z}_j]$ denotes the $n\times n$ matrix of commutators (which are scalars).}

Just like Eq. (\ref{birulaEUR}) can be extended to Eq. (\ref{EURrenyi}), the entropic uncertainty relation (\ref{EURFRFTcom}) can also be extended to Rényi entropies: 

\theorem{Rényi EUR for two arbitrary vectors of intercommuting quadratures \cite{hertz3}}{
		Let $\mathbf y = (\hat y_1,\cdots \hat y_n)^T$ be a vector of commuting quadratures, $\mathbf z = (\hat z_1,\cdots \hat z_n)^T$ be another vector of commuting quadratures, and let each of the components of these vectors be written as a linear combination of the $(\hat x, \hat p)$ quadratures of an $N$-modal system ($N \ge n$).
		Then, any $N$-modal state $\rho$ verifies the Rényi entropic uncertainty relation 
		\begin{eqnarray}
		h_\alpha(\mathbf y)+h_\beta(\mathbf z)&\geq&n\ln(\pi)+\frac{n\ln(\alpha)}{2\left(\alpha-1\right)}+\frac{n\ln(\beta)}{2\left(\beta-1\right)} +\ln\left|\det \mathbf{K}\right|.
		\label{renyiEURFRFT}
		\end{eqnarray}
		with
		\begin{equation}
		\frac{1}{\alpha}+\frac{1}{\beta}=2,  \qquad \alpha>0, \qquad \beta>0,
		\label{alphabeta}
		\end{equation}
		where $h_\alpha(\cdot)$ stands for the Rényi entropy of the probability distributions of the vectors of jointly measured quadratures $\hat y_i$'s or $\hat z_j$'s, and $\mathbf{K}_{ij}=[\hat y_i,\hat z_j]$ is the $n\times n$ matrix of commutators (which are scalars).
}

	As expected, in the limit where $\alpha\rightarrow1$ and $\beta\rightarrow1$, we recover the uncertainty relations for Shannon differential entropies, Eq. (\ref{EURFRFTcom}). Moreover, in the one-dimensional case ($N=~n=~1$), Eq. (\ref{renyiEURFRFT}) coincides with the result found in \cite{Guanlei2}.
	
	\subsection{Other entropic uncertainty relations}
\label{otherEUR}

 Let us conclude this section by mentioning a few related entropic uncertainty relations. First, it is also possible to express with entropies the complementary between the pair of variables $(\phi,L_z)$, that is, a (continuous) angle and associated (discrete) angular momentum \cite{birula1984}, or $(\phi, \hat{N})$, that is, a (continuous) phase and associated (discrete) number operator \cite{Hall1993, Hall}. Unlike those considered in the present paper, such entropic uncertainty relations for $(\phi,L_z)$ or $(\phi, \hat{N})$ may be viewed as hybrid as they mix discrete and continuous entropies. Similarly, Hall considered an entropic time-energy uncertainty relation for bound quantum systems (thus having discrete energy eigenvalues), which expresses the balance between a discrete entropy for the energy distribution and a continuous entropy for the time shift applied to the system \cite{Hall2008}. Recently, an entropic time-energy uncertainty relation has also been formulated for general time-independent Hamiltonians, where the time uncertainty is associated with measuring the (continuous) time state of a quantum clock \cite{Coles2018}.

As already mentioned, another variant of entropic uncertainty relations can be defined in the presence of quantum memory. This situation, where the observer may exploit some side information, has been analyzed in the case of position and momentum variables by Furrer {\it et al.} \cite{Furrer2014}. Still another interesting scenario concerns uncertainties occurring in successive measurements. While the most common uncertainty relations assume that one repeats measurements on the same state (as we do throughout this paper), one may consider successive measurements on a system whose state evolves as a result of the measurements. The entropic uncertainty relations for canonically conjugate variables in this scenario have been derived by Rastegin \cite{rastegin2016}.

This closes our review on continuous-variable entropic uncertainty relations. In the rest of this paper, we present some new results, namely a tight entropic uncertainty relation for two vectors of quadratures (Section~\ref{2EUR}) and several conjectures for new entropic uncertainty relations involving more than two variables (Section~\ref{secConj}).


\section{Tight entropic uncertainty relation for arbitrary quadratures} 

\label{2EUR} 

\subsection{Minimum uncertainty states}

We now build an entropic uncertainty relation which holds for any two vectors of  intercommuting quadratures and is saturated by all pure Gaussian states (hence, we call it tight).
It combines the two previous results, namely Eqs. (\ref{nmodes}) and (\ref{EURFRFTcom}). 

\begin{figure}[h]
	\centering	\includegraphics[width=0.6\columnwidth]{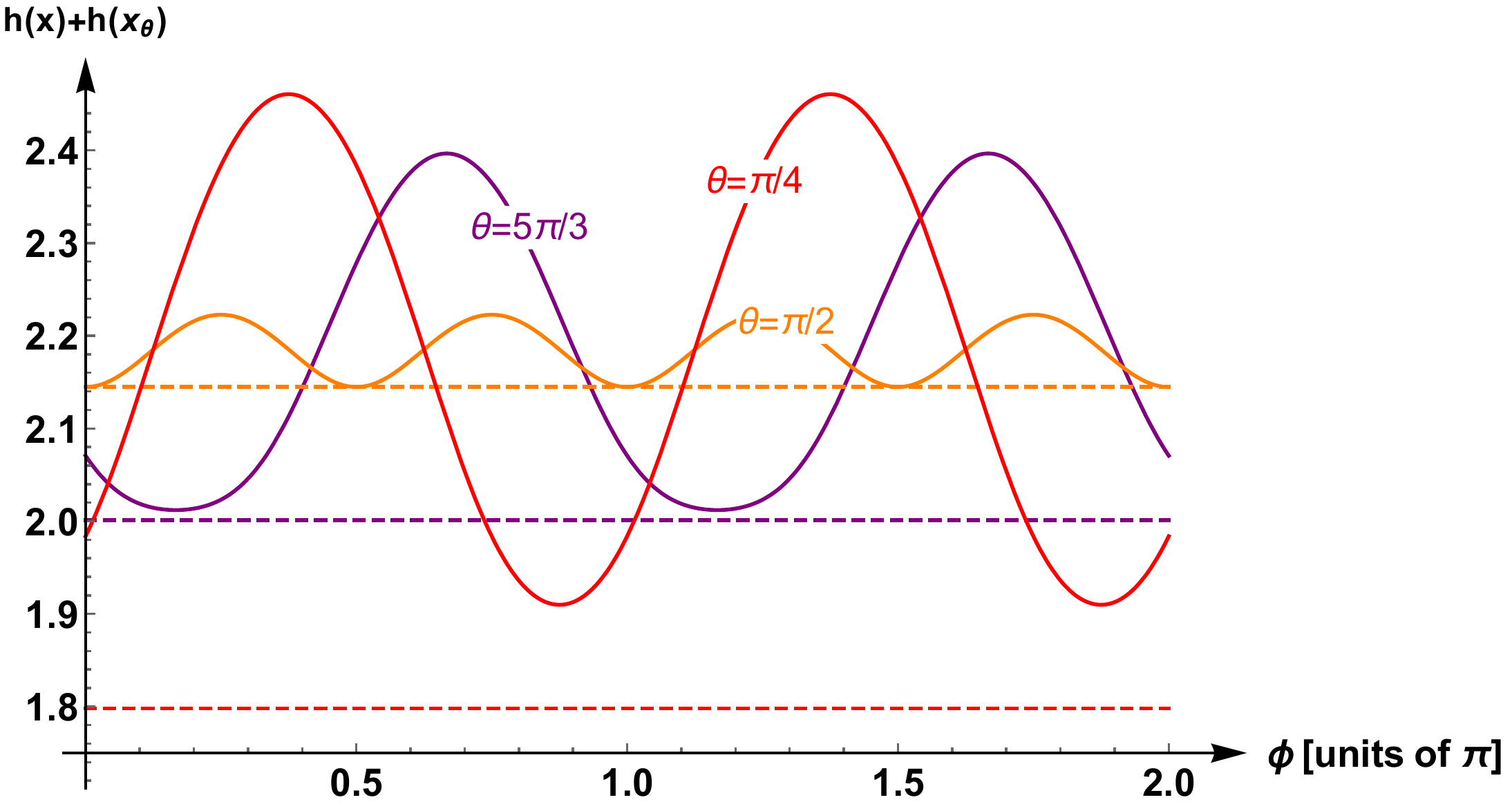}
	\caption{\label{saturation_guanlei} Plot of Eq.~(\ref{EQunmode}) for pure Gaussian states, illustrating that this entropic uncertainty relation is, in general, not saturated. Solid lines represent the sum of entropies for $\theta=\pi/4,\pi/2 \text{ and }5\pi/3$, while dashed lines show the corresponding lower bounds $\ln(\pi e |\sin\theta|)$. }
\end{figure}

Let us stress that Eq.~(\ref{EURFRFTcom}) is not saturated by pure Gaussian states, in general, so the idea here is to take correlations into account following a similar procedure as the one leading to Eq.~(\ref{nmodes}). One can easily understand the problem by considering the one mode case, Eq.~(\ref{eq-Guanlei}), which can also be written as
\eq{h(x)+h(x_\theta)\geq\ln(\pi e|\sin\theta|).\label{EQunmode}}
We compute $h(x)+h(x_\theta)$ for a general pure Gaussian state (i.e., a squeezed state with parameter $r$ and angle $\phi$) with covariance matrix
\eq{\gamma=	\frac{1}{2} \left(
\begin{array}{cc}
e^{-2 r} \cos ^2\phi +e^{2 r} \sin ^2\phi  &  (e^{2 r} -e^{-2 r})\cos \phi  \sin \phi   \\
 (e^{2 r} -e^{-2 r})\cos \phi  \sin \phi  &e^{2 r} \cos ^2\phi +e^{-2 r} \sin ^2\phi  \\
\end{array}
\right)\label{gammaSState}
}
and plot it as a function of $\phi$ (we fix $r=0.2$ and consider several values of $\theta$).
As we can see on Figure~\ref{saturation_guanlei}, where the solid lines represent the sum of entropies (for $\theta=\pi/4,\pi/2 \text{ and }5\pi/3$) and the dashed lines stands for the corresponding lower bound $\ln(\pi e |\sin\theta|)$, the uncertainty relation (\ref{EQunmode}) is in general not saturated by any pure Gaussian state. The only exception is $\theta=\pi/2$, namely the Białynicki-Birula and Mycielski uncertainty relation (\ref{birulabis}), which is only saturated when the state is aligned with the principal axes, i.e. if $\phi=0,\pi/2,\pi \text{ or }3\pi/2$. 

This suggests that a modification of Eq. (\ref{EURFRFTcom}) is needed in order to impose that the Gaussian pure states become minimum-uncertainty states, as they are in Eq.  (\ref{nmodes}).

\subsection{Entropic uncertainty relation saturated by all pure Gaussian states}

\label{defRelation}
Let $\mathbf y = (\hat y_1,\cdots \hat y_n)^T$  be a vector of commuting quadratures and $\mathbf z =~(\hat z_1,\cdots \hat z_n)^T$ be another vector of commuting quadratures. Let us suppose that they correspond to the output $x$-quadratures obtained after applying two possible symplectic transformations onto some $n$-modal system. In other words, the $2n$-dimensional vector of input quadratures $(\hat x_1,\cdots ,\hat x_n,\hat p_1,\cdots ,\hat p_n)^T$ is transformed into the $2n$-dimensional vector of output quadratures $(\hat y_1,\cdots ,\hat y_n,\hat q_1,\cdots ,\hat q_n)^T$ or $(\hat z_1,\cdots ,\hat z_n,\hat o_1,\cdots ,\hat o_n)^T$,
where $(\hat q_1,\cdots ,\hat q_n)^T$ [resp.  $(\hat o_1,\cdots ,\hat o_n)^T$] is a vector of quadratures that are pairwise canonically conjugate with $(\hat y_1,\cdots ,\hat y_n)^T$ [resp. $(\hat z_1,\cdots ,\hat z_n)^T$]. 
As  for the $x,p$ quadratures, it is possible to define a covariance matrix $\mathbf{\Gamma}$ for the  $y,z$ quadratures. Its elements are expressed as
\begin{equation}
	\mathbf{\Gamma}_{ij}={1\over 2}\mean{\hat R_i\hat R_j+\hat R_j\hat R_i}-\mean{\hat R_i}\mean{\hat R_j}
	\label{gammayz}
\end{equation}
with ${\bf R}=(\hat y_1,...,\hat y_n,\hat z_1,...,\hat z_n)^T$.
The knowledge of $\mathbf{\Gamma}$ allows us to take correlations into account and write a general form of the entropic uncertainty relation for any two vectors of intercommuting quadratures:

\theorem{Tight EUR for two arbitrary vectors of intercommuting quadrature}
	{Let $\mathbf y =~(\hat y_1,\cdots \hat y_n)^T$ be a vector of commuting quadratures, $\mathbf z =~(\hat z_1,\cdots \hat z_n)^T$ be another vector of commuting quadratures, and let each of the components of $\mathbf y$ and $\mathbf z$ be written as a linear combination of the $(\hat x, \hat p)$ quadratures of an $n$-modal system. Then, any $n$-modal state $\rho$ satisfies the  entropic uncertainty relation \footnote{The proof is conditional on two reasonable assumptions, see below.}
	\begin{equation}
	h({\bf y})+h({\bf z})-\frac{1}{2}\ln\left(\frac{\det\mathbf{\Gamma}_{\mathbf{y}}\det\mathbf{\Gamma}_{\mathbf{z}}}{\det\mathbf{\Gamma}}\right)\geq\ln\left((\pi e)^n | \det \mathbf{K}|\right)
	\label{NEWEUR}
\end{equation}
where  $h(\cdot)$ stands for the Shannon differential entropy of the probability distribution of the vector of jointly measured quadratures $\hat y_i$'s or $\hat z_j$'s, 
$\mathbf{\Gamma}$ is the covariance matrix defined in Eq. (\ref{gammayz}), $\mathbf{\Gamma}_{\mathbf{y}}$ and $\mathbf{\Gamma}_{\mathbf{z}}$ are the reduced covariance matrices of the $\hat y_i$ and $\hat z_i$ quadratures, respectively, and $\mathbf{K}_{ij}=[\hat y_i,\hat z_j]$ is the $n\times n$ matrix of commutators (which are scalars). The saturation is obtained when $\rho$ is Gaussian and pure.
}

 Let us first remark that Eq.~(\ref{NEWEUR}) is invariant under displacements. Indeed, the differential entropy is invariant under displacements [see Eq. (\ref{translation-h})], and so are $\mathbf{\Gamma}$, $\mathbf{\Gamma}_{\mathbf{y}}$ and $\mathbf{\Gamma}_{\mathbf{z}}$ as is obvious from their definitions. Thus, in the proof of Eq.~(\ref{NEWEUR}), we can restrict to states centered at the origin.
As we will see in Section \ref{proofEURgen}, our proof is based on a variational method used to show that pure Gaussian states extremize the uncertainty functional
\begin{equation}
	F(\hat \rho) = h({\bf y})+h({\bf z})   -  {1\over 2} \ln \left(\frac{\det \mathbf{\Gamma}_{\mathbf{y}}\det \mathbf{\Gamma}_{\mathbf{z}}}{\det \mathbf{\Gamma}}\right) .
	\label{uncertainty-functional-n-modebis}
\end{equation}
The proof, however, is partial as it relies on two assumptions:
\begin{enumerate}
\item \label{Assum1bis}Pure Gaussian states are global minimizers of the uncertainty functional $F(\hat \rho)$.
\item  \label{Assum2bis}The uncertainty functional $F(\hat \rho)$ is concave, so relation (\ref{NEWEUR}) is  valid.
\end{enumerate} 

\subsection{Special case of Gaussian states}
\label{exGS}
Before addressing the proof of Eq.~(\ref{NEWEUR}) with a variational method, let us see how this entropic uncertainty relation applies to Gaussian states. In particular, let us prove first that Eq.~(\ref{NEWEUR}) is saturated by all {\it pure} Gaussian states. Then, we will show that for all Gaussian states, it can be proven using the $n$-modal version of  the Robertson-Schrödinger uncertainty relation, Eq. (\ref{RSanmodes}).

Consider a {\it pure} $n$-modal Gaussian state. Its Wigner function  is given by 
\begin{equation}
	W^G(x,p)=\frac{1}{\pi^n}e^{-{1\over2}{\bf r}^T\gamma^{-1}{\bf r} }
\end{equation}
and its covariance matrix is expressed as
\begin{equation}
	\gamma=\begin{pmatrix}
		\gamma_{\mathbf{x}}&\gamma_{\mathbf{xp}}\\\gamma_{\mathbf{xp}}&\gamma_{\mathbf{p}}
	\end{pmatrix}_{2n\times 2n}
\end{equation}
where $\gamma_{\mathbf{x}}$ and $\gamma_{\mathbf{p}}$ are the reduced covariance matrices of the position and momentum quadratures. Since the state is pure and Gaussian, $\det\gamma=(1/4)^n$.

To evaluate Eq. (\ref{NEWEUR}) we need to find the determinant of the covariance matrix $\mathbf{\Gamma}$ for the $y,z$-quadratures. The calculation is reported in \ref{appB} and leads to \begin{equation}
	\det \mathbf{\Gamma}=\det \gamma\, | \det \mathbf{K}|^2.
	\label{equivgammaK}
\end{equation} 
Note that Eq. (\ref{equivgammaK}) is true for any state, Gaussian or not, but since we are dealing with a pure Gaussian state,  $\det \gamma=1/ 4^{n}$, it simplifies to
\begin{equation}
	\det \mathbf{\Gamma}=\frac{1}{4^n}|\det \mathbf{K}|^2.
	\label{gammaGetK}
\end{equation}
The last step needed to evaluate Eq.~(\ref{NEWEUR}) is to compute the differential entropies of the $y$ and $z$ quadratures. Since these quadratures are obtained after applying some symplectic transformations, the Wigner function, which is Gaussian for the input state, remains Gaussian for the output state. The probability distributions of the jointly measured quadratures $\hat y_i$ or $\hat z_j$ are thus given by the following Gaussian distributions
\begin{equation}
	P({\bf y})=\frac{1}{\sqrt{(2\pi)^n\det \mathbf{\Gamma}_{\mathbf{y}}}}e^{-{1\over2}{\bf y}^T\mathbf{\Gamma}_{\mathbf{y}}^{-1}{\bf y} },\qquad\qquad
	P({\bf z})=\frac{1}{\sqrt{(2\pi)^n\det \mathbf{\Gamma}_{\mathbf{z}}}}e^{-{1\over2}{\bf z}^T\mathbf{\Gamma}_{\mathbf{z}}^{-1}{\bf z}}
	\label{Pgauss}
\end{equation}
and we easily evaluate the corresponding differential entropies
\begin{equation}
	h({\bf y})=\frac{1}{2}\ln \big((2\pi e)^n\det \mathbf{\Gamma}_{\mathbf{y}}\big),\qquad\qquad
	h({\bf z})=\frac{1}{2}\ln \big((2\pi e)^n\det \mathbf{\Gamma}_{\mathbf{z}}\big )\label{entro}.
\end{equation}
Inserting these quantities together with Eq.~(\ref{gammaGetK}) into the left-hand side of Eq.~(\ref{NEWEUR}) yields the lower bound $\ln((\pi e)^n|\det \mathbf{K}|)$, so we have proved that Gaussian pure states are minimum uncertainty states of Eq.~(\ref{NEWEUR}), as desired.

Let us emphasize that the entropic uncertainty relation that does not take correlations into account, Eq.~(\ref{EURFRFTcom}), is only  saturated by pure Gaussian states with vanishing correlations. Indeed, for pure Gaussian states, we find that
\eq{h({\bf y})+h({\bf z})=\ln \big((2 \pi e)^{n}\sqrt{\det \mathbf{\Gamma}_{\mathbf{y}}\det \mathbf{\Gamma}_{\mathbf{z}}}\big) }
reaches the lower bound of Eq. (\ref{EURFRFTcom}) only if 
\eqarray{2 ^{n}\sqrt{\det \mathbf{\Gamma}_{\mathbf{y}}\det \mathbf{\Gamma}_{\mathbf{z}}}&=&|\det \mathbf{K}|\nonumber\\
	\Leftrightarrow\quad2 ^{n}\sqrt{\det \mathbf{\Gamma}_{\mathbf{y}}\det \mathbf{\Gamma}_{\mathbf{z}}}&=&2^n \sqrt{\det \mathbf{\Gamma}} \nonumber\\
	\Leftrightarrow\qquad\quad\det \mathbf{\Gamma}_{\mathbf{y}}\det \mathbf{\Gamma}_{\mathbf{z}}&=&\det \mathbf{\Gamma}
}
where we have used Eq. (\ref{gammaGetK}). Obviously, this is only true when $\mathbf{\Gamma}_{yz}=0$, i.e., when there is no correlation between the $y_i$ and $z_i$ quadratures. This confirms that the entropic uncertainty relation (\ref{EURFRFTcom}) is not saturated by all pure Gaussian states, as we had explicitly checked for one mode in Figure~\ref{saturation_guanlei}.

Second, let us now prove that Eq.~(\ref{NEWEUR}) holds  for a general mixed Gaussian state.
For any Gaussian state, pure or not, the differential entropies are still given by Eq.~(\ref{entro}), so that
	\eq{h(\mathbf{y})+h(\mathbf{z})-\frac{1}{2}\left(\frac{\det\mathbf{\Gamma}_{\mathbf{y}}\det\mathbf{\Gamma}_{\mathbf{z}}}{\det\mathbf{\Gamma}}\right)=\ln \left((2\pi e)^n\sqrt{\det\mathbf{\Gamma}}\right).
		\label{EURgenforGS}	}
	Using Eq.~(\ref{equivgammaK}) together with the $n$-modal version of the Robertson-Schrödinger uncertainty relation, $\det\gamma\geq1/4^n$, we get
	\eq{\sqrt{\det\mathbf{\Gamma}}\geq\frac{|\det \mathbf{K}|}{2^n}}
	Injecting this inequality into Eq. (\ref{EURgenforGS}) complete the proof of Eq.~(\ref{NEWEUR}) for all Gaussian states.
	


\subsection{Partial proof for all states}
\label{proofEURgen}

The difficult part is to verify the entropic uncertainty relation for a general -- not necessarily Gaussian -- state. Inspired from \cite{hertz2}, we give here a partial proof of Eq.~(\ref{NEWEUR}) based on a variational method (see Ref.~\cite{jackiw, Hall}), which is conditional on two assumptions [see Assumptions (\ref{Assum1bis}) and (\ref{Assum2bis}) in Section \ref{defRelation}]. More precisely, we seek a pure state $\hat \rho=\ket{\psi}\bra{\psi}$ that extremizes our uncertainty functional (\ref{uncertainty-functional-n-modebis})
and show that any pure Gaussian state is such an extremum.
The steps of the proof are  similar to those developed in \cite{hertz2}, except that we consider the $y,z$-quadratures instead of the $x,p$-quadratures. The assumptions are also the same.

	As already mentioned, $ F(\ket{\psi})$ is invariant under displacements so that we can restrict our search to extremal states centered on $0$. We also require extremal state to be normalized. Accounting for these constraints by using the Lagrange multipliers method, we have to solve ${{\partial J}\over{\partial \bra{\psi}}}=0$ with
	\begin{equation}
		J=h({\bf y})+h({\bf z})   -  {1\over 2} \ln \left(\frac{\det \mathbf{\Gamma}_{\mathbf{y}}\det \mathbf{\Gamma}_{\mathbf{z}}}{\det \mathbf{\Gamma}} \right)+\lambda( \mean{\psi|\psi}-1) +\sum_{i=1}^{2n}\mu_i \mean{\psi|\hat{R_i}|\psi} ,
         \label{J-Lagrange}
	\end{equation}
where $\lambda$ and $\mu_i$  are Lagrange multipliers.  Note that, as  explained in \cite{hertz2}, it is not necessary to consider ${{\partial J}\over{\partial \ket{\psi}}}=0$ since no  additional information would be obtained.
	
Let us evaluate the derivative of each term of Eq.~(\ref{J-Lagrange}) separately. First, the derivative of $h({\bf y})$ gives
	\begin{eqnarray}   \label{function-of-operator}
		\frac{\partial h({\bf y})}{\partial \bra{\psi}}&=& \frac{\partial }{\partial \bra{\psi}}\left(\int P({\bf y})\ln P({\bf y})d{\bf y}\right)\nonumber\\
		&=&\frac{\partial }{\partial \bra{\psi}}\left(\int \mean{\psi|{\bf y}}\mean{{\bf y}|\psi}\ln( \mean{\psi|{\bf y}}\mean{{\bf y}|\psi})d{\bf y}\right)\nonumber\\
		&=&-\left(  \ln P({\bf y}) +1\right) \ket{\psi} 
	\end{eqnarray}
	and similarly for $h({\bf z})$. Note that ${\bf y}$ in the last line of Eq.~(\ref{function-of-operator}) denotes a vector of quadrature operators, so that $\ln P({\bf y})$ is an operator too.
	With the help of Jacobi's formula \cite{Magnus},  the derivatives of the determinant of the three covariance matrices give
	\begin{eqnarray}
		\frac{\partial }{\partial \bra{\psi}}\ln\det \mathbf{\Gamma}_{\mathbf{y}}&=&\frac{1}{\det \mathbf{\Gamma}_{\mathbf{y}}}   \frac{\partial }{\partial \bra{\psi}}\det \mathbf{\Gamma}_{\mathbf{y}} \nonumber\\
		&=&\frac{1}{\det \mathbf{\Gamma}_{\mathbf{y}}} \mathrm{Tr}\left[(\det \mathbf{\Gamma}_{\mathbf{y}})\,\mathbf{\Gamma}_{\mathbf{y}}^{-1}\frac{\partial \mathbf{\Gamma}_{\mathbf{y}}}{\partial \bra{\psi}}\right] \nonumber\\
		&=&\sum_{i=1}^{n}\sum_{k=1}^{n}(\mathbf{\Gamma}_{\mathbf{y}})_{ik}^{-1}\frac{\partial(\mathbf{\Gamma}_{\mathbf{y}})_{ki}}{\partial\bra{\psi}}\nonumber\\
		&=&\sum_{i=1}^{n}\sum_{k=1}^{n}(\mathbf{\Gamma}_{\mathbf{y}})_{ik}^{-1}\frac{(\hat{y}_k\hat{y}_i+\hat{y}_i\hat{y}_k)}{2}\ket{\psi}\nonumber\\
		&=&\left[\sum_{i=1}^{n}\sum_{k=1}^{n}\frac{\hat{y}_k (\mathbf{\Gamma}_{\mathbf{y}})_{ik}^{-1}\hat{y}_i}{2}+\sum_{i=1}^{n}\sum_{k=1}^{n}\frac{\hat{y}_i (\mathbf{\Gamma}_{\mathbf{y}})_{ik}^{-1}\hat{y}_k}{2} \right]\ket{\psi} \nonumber\\
		&=&{\bf y}^T\mathbf{\Gamma}_{\mathbf{y}}^{-1}{\bf y}\,\ket{\psi}.
	\end{eqnarray}
	and similarly
	\begin{eqnarray}
		\frac{\partial }{\partial \bra{\psi}}\ln\det \mathbf{\Gamma}_{\mathbf{z}}
		&=&{\bf z}^T\mathbf{\Gamma}_{\mathbf{z}}^{-1}{\bf z}\,\ket{\psi}\nonumber\\
		\frac{\partial }{\partial \bra{\psi}}\ln\det \mathbf{\Gamma}
		&=&{\bf R}^T\mathbf{\Gamma}^{-1}{\bf R}\,\ket{\psi}.
	\end{eqnarray}
	Finally, the derivative of the last two terms of Eq.~(\ref{J-Lagrange}) give
	\begin{eqnarray}
		\frac{\partial}{\partial \bra{\psi}} \bigg(  \lambda( \mean{\psi|\psi}-1) + \sum_{i=1}^{2n}\mu_i \mean{\psi|\hat{R}_i|\psi}  \bigg)    
		= \left(\lambda+\sum_{i=1}^{2n}\mu_i \hat{R}_i\right)\ket{\psi}
	\end{eqnarray}
	so that the variational equation can be rewritten as an eigenvalue equation for $ \ket{\psi}$,
	\begin{eqnarray}  \label{eigenequ2}
		\bigg[-\ln P({\bf y}) -\ln P({\bf z})  -2+\lambda&+&\sum_{i=1}^{2n}\mu_i \hat{R}_i 
		- {1\over 2}{\bf y}^T\mathbf{\Gamma}_{\mathbf{y}}^{-1}{\bf y}  \nonumber \\
		&-& {1\over 2}{\bf z}^T\mathbf{\Gamma}_{\mathbf{z}}^{-1}{\bf y} \,+ \,{1\over 2}{\bf R}^T\mathbf{\Gamma}^{-1}{\bf R}\,\,
		\bigg]  \ket{\psi}=0 . 
	\end{eqnarray}
	Thus, the states $\ket{\psi}$ extremizing $ F(\ket{\psi})$ are the eigenstates of Eq. (\ref{eigenequ2}).

	Now, instead of looking for all  eigenstates, we show that all pure Gaussian states are  solution of Eq.~(\ref{eigenequ2}).
	We have already written the probability distributions $P({\bf y})$ and $P({\bf z})$ for a $n$-modal pure Gaussian state [see Eq.~(\ref{Pgauss})], 
	so we have
	\begin{equation}
		\ln P({\bf y}) +\ln P({\bf z})  =-\ln\left((2\pi)^n \sqrt{\det\mathbf{\Gamma}_{\mathbf{y}}\det\mathbf{\Gamma}_{\mathbf{z}}} \right)
		-  {1\over 2}{\bf y} ^T\mathbf{\Gamma}_{\mathbf{y}}^{-1}{\bf y} -  {1\over 2}{\bf z} ^T\mathbf{\Gamma}_{\mathbf{z}}^{-1}{\bf z} 
		\label{sumW}
	\end{equation}
	and the eigenvalue equation (\ref{eigenequ2}) reduces to
	\begin{equation}
		\bigg[\ln\left((2\pi)^n \sqrt{\det\mathbf{\Gamma}_{\mathbf{y}}\det\mathbf{\Gamma}_{\mathbf{z}}} \right) -2+\lambda+\sum_{i=1}^{2n}\mu_i \hat{R}_i     + \,{1\over 2}{\bf R}^T\mathbf{\Gamma}^{-1}{\bf R}\,\,
		\bigg]  \ket{\psi}=0 . 
		\label{eigenequ3}
	\end{equation}
As shown in \ref{appC}, pure $n$-modal Gaussian states (centered on the origin) are eigenvectors of 
	 ${1\over 2} \,{\bf R} ^T\mathbf{\Gamma}^{-1}{\bf R}$ with eigenvalue $n$, that is
	 \begin{equation}
{1\over 2} \,{\bf R} ^T\mathbf{\Gamma}^{-1}{\bf R}\ket{\psi}=n \, \ket{\psi} ,
          \label{eigenequ3bis}
	 \end{equation}
so that Eq. (\ref{eigenequ3}) can be further simplified to
	\begin{eqnarray}
		\bigg[\ln\left((2\pi)^n \sqrt{\det\mathbf{\Gamma}_{\mathbf{y}}\det\mathbf{\Gamma}_{\mathbf{z}}} \right) +n-2+\lambda+\sum_{i=1}^{2n}\mu_i \hat{R}_i   
		\bigg]  \ket{\psi}=0 . 
		\label{eigenequ4}
	\end{eqnarray}
	The value of $\lambda$ is found by multiplying this equation on the left by $\bra{\psi}$ and using the normalization constraint $ \mean{\psi|\psi}=1$, as well as the fact that the mean values vanish, $\mean{\psi|\hat{R}_i|\psi}=0$ for all $i$, so that we are left with 
	\begin{equation}
		\left[\sum_{i=1}^{2n}\mu_i\hat{R}_i\right]\ket{\psi}=0
	\end{equation}
	which is satisfied if we set all the $\mu_i=0$.
	
	In summary, we have proved that there exists an appropriate choice for $\lambda$ and $\mu_i$ such that any pure Gaussian state centered on the origin is an extremum of the uncertainty functional $F(\ket{\psi})$, that is, any $n$-modal squeezed vacuum state (with arbitrary squeezing and orientation) extremizes $F(\ket{\psi})$. Since this functional is invariant under displacement, this feature extends to all pure Gaussian states. According to Assumption (\ref{Assum1bis}), we take for granted that pure Gaussian states are not just local extrema, but global minima of the uncertainty functional. The last step is simply to evaluate the functional for Gaussian pure states and see that it yields $\ln((\pi e)^n|\det \mathbf{K}|)$, as shown in Section \ref{exGS}. This  completes the proof of Eq. (\ref{NEWEUR}) for pure states. To complete the proof for mixed states, we resort to Assumption (\ref{Assum2bis}): if the functional $F(\hat\rho)$ is concave and Eq.~(\ref{NEWEUR}) holds for pure state, then it is necessarily true for mixed states too.

\subsection{Alternative formulation}

Interestingly, using the relation between $\det\mathbf{\Gamma}$ and $\det \mathbf{K}$ exhibited by Eq.~(\ref{equivgammaK}), we can rewrite our tight entropic uncertainty relation (\ref{NEWEUR}) without the explicit dependence on the commutator matrix $\mathbf{K}$, that is
\begin{equation}
h({\bf y})+h({\bf z})-\frac{1}{2}\ln\left(\frac{\det \mathbf{\Gamma}_{\mathbf{y}}\det \mathbf{\Gamma}_{\mathbf{z}}}{\det \gamma}\right)\geq\ln\left((\pi e)^n \right).
\label{Alternative-tight-EUR}
\end{equation}
Here $\gamma$ is the covariance matrix for the $x,p$-quadratures, while $\mathbf{\Gamma}_{\mathbf{y}}$ and $\mathbf{\Gamma}_{\mathbf{z}}$ are the reduced covariance matrices of the $y,z$ quadratures. If we know $\gamma$ and the symplectic transformations leading to $\mathbf{y}$ and $\mathbf{z}$, it is straightforward to access  $\mathbf{\Gamma}_{\mathbf{y}}$ and $\mathbf{\Gamma}_{\mathbf{z}}$ through Eq.~(\ref{Gammagamma}), which makes the computation of Eq.~(\ref{Alternative-tight-EUR}) easier. Note also that this alternative formulation becomes very similar to the tight entropic uncertainty relation for canonically-conjugate variables $\mathbf{x}$ and $\mathbf{p}$, Eq. (\ref{nmodes}), where we simply substitute $\mathbf{\Gamma}_{\mathbf{y}}$ for $\gamma_{\mathbf{x}}$ and $\mathbf{\Gamma}_{\mathbf{z}}$ for $\gamma_{\mathbf{p}}$.

\subsection{Entropy-power formulation and covariance-based uncertainty relation}
\label{entrop-pow-formul-of-tight-EUR}

Following the same procedure as before, we may exploit the entropy-power formulation in order to rewrite Eq. (\ref{NEWEUR}) as
\begin{equation}
	N_{\mathbf{y}} \, N_{\mathbf{z}} \,  \left(\frac{\det \mathbf{\Gamma}}{\det\mathbf{\Gamma}_{\mathbf{y}}\det\mathbf{\Gamma}_{\mathbf{z}}}\right)^{1/n}  \geq \frac{ |\det \mathbf{K}|^{2/n} }{4}  .
	\label{entropy-power-ur2}
\end{equation} 
which is a tight entropy-power uncertainty relation for two arbitrary vectors of quadratures $\mathbf{y}$ and $\mathbf{z}$. This entropy-power formulation helps us better see that the tight entropic uncertainty relation Eq. (\ref{NEWEUR}) implies  Eq.~(\ref{EURFRFTcom}). Indeed, since $\det\mathbf{\Gamma}_{\mathbf{y}}\det\mathbf{\Gamma}_{\mathbf{z}}\geq\det\mathbf{\Gamma},$\footnote{This is a generalization of Hadamard's inequality, Eq.~(\ref{hadam})} we see that Eq.~(\ref{entropy-power-ur2}) corresponds to lifting up the lower bound on $N_{\mathbf{y}} \, N_{\mathbf{z}}$ in Eq.~(\ref{entropy-power-ur}) by a term that accounts for the $\mathbf{y},\mathbf{z}$ correlations. Thus Eq.~(\ref{entropy-power-ur2}) implies Eq.~(\ref{entropy-power-ur}), which is the entropy-power version of Eq.~(\ref{EURFRFTcom}).

Now, we again use the fact that the maximum entropy for a fixed covariance matrix is reached by the Gaussian distribution, so that we can upper bound $N_{\mathbf{y}}N_{\mathbf{z}}$ by $(\det\mathbf{\Gamma}_{\mathbf{y}}\det\mathbf{\Gamma}_{\mathbf{z}})^{1/n}$. Combining this with Eq.~(\ref{entropy-power-ur2}), we obtain the variance-based uncertainty relation for two arbitrary vectors of quadratures $\mathbf{y}$ and $\mathbf{z}$,
\eq{   \det \mathbf{\Gamma}   \geq \frac{ |\det \mathbf{K}|^2 }{4^n}
	\label{RobGen}}
which generalizes Eq. (\ref{URcovariance}) as it takes the $\mathbf{y},\mathbf{z}$ correlations into account.

Interestingly, Eq. (\ref{RobGen}) is nothing else but a special case of the Robertson uncertainty relation (\ref{robGen}). Indeed, we see that the definition of $\mathbf \Gamma$ in Eq. (\ref{gammayz}) coincides with that of Eq. (\ref{sigmaC}), with $m=2n$. Here, we have $R_i=y_i$ and $R_{n+i}=z_i$  for $i=1,\cdots,n$, so that the matrix $\mathbf C$ defined in  Eq. (\ref{sigmaC}) can be written in terms of $\mathbf{K}_{ij}=[\hat y_i,\hat z_j]$  as
\eq{\mathbf C=-\frac{i}{2}\begin{pmatrix}
		0_{n\times n}&\mathbf{K}\\-\mathbf{K}&0_{n\times n}
		\label{Cdiag}
\end{pmatrix}}
Therefore, 
\eq{\det \mathbf C=\left(-\frac{i}{2}\right)^{2n}(\det \mathbf{K})^2=\frac{ |\det \mathbf{K}|^2}{4^n} }
where we used the fact that the $\mathbf{K}_{ij}$'s are all pure imaginary numbers, implying that Eq. (\ref{robGen}) reduces to Eq. (\ref{RobGen}) in this case.


\section{Entropic uncertainty relations for more than two observables}

\label{secConj} 

All entropic uncertainty relations considered in Sections \ref{EURintro} and \ref{2EUR} address the case of two variables (or two vectors consisting each of $n$ commuting variables). Here, we turn to entropic uncertainty relations for more than two variables. As already mentioned, the entropy-power formulation is convenient to show that, in general, an entropic uncertainty relation implies a variance-based one. In particular, we showed in Secion~\ref{entrop-pow-formul-of-tight-EUR} that Eq.~(\ref{NEWEUR}) implies the Robertson uncertainty relation, Eq.~(\ref{robGen}). More precisely, we have shown that it only implies a special case of it, namely Eq. (\ref{RobGen}). Therefore, it is natural to conjecture that there exists a more general entropic uncertainty relation which implies the Robertson uncertainty relation for any matrix ${\mathbf C}$, more general than in Eq.~(\ref{Cdiag}). 
Our first conjecture is an extension of Eq.~(\ref{NEWEUR}) for $m$ variables:
\begin{conj}  \label{conj1}
	Any $n$-modal state $\rho$ satisfies the entropic uncertainty relation
	\eq{h(R_1)+h(R_2)+\cdots+h(R_m)-\frac{1}{2}\ln\left(\frac{\sigma_1^2\sigma_2^2\cdots\sigma_m^2}{\det\mathbf\Gamma}\right)\geq\frac{1}{2}\ln((2\pi e)^m\det \mathbf C)\label{conjGenEUR}}
	where $\hat R_i$'s are $m$ arbitrary continuous observables, $\sigma_i^2$ is the variance of each $\hat R_i$ while $\mathbf\Gamma$ is the covariance matrix of the $\hat R_i$'s, and $\mathbf C$ is the matrix of commutators. The elements of $\mathbf\Gamma$ and $\mathbf C$ are defined as in Eq. (\ref{sigmaC}).
\end{conj}

This entropic uncertainty relation is valid regardless of whether the $\hat R_i$'s commute or not, but is interesting for an even number of them only. Indeed, as mentioned for the Robertson uncertainty relation, Eq.~(\ref{robGen}), when $m$ is odd, $\det \mathbf C=0$ and the lower bound in Conjecture \ref{conj1} equals $-\infty$. Note also that Eq.~(\ref{conjGenEUR}) is defined only when $m\leq 2n$, where $n$ is the number of modes of the state $\rho$. Indeed, if $m>2n$, the determinant of the covariance matrix $\mathbf\Gamma$ vanishes (we can always write one column as a linear combination of two other columns). This is consistent with the fact that  $\det \mathbf C$ vanishes in this case too. Indeed, $\det \mathbf C \ge 0$ since $\mathbf C$ is an anti-symmetric matrix \cite{Cayley}, so that Eq.~(\ref{robGen}) implies that if $\det \mathbf \Gamma$ is null, so is $\det \mathbf C$.

Finally, let us mention that Eq.~(\ref{conjGenEUR}) is invariant under the scaling of one variable. Assume, with no loss of generality, that $R_1\rightarrow R_1'=a R_1$, where $a$ is some scaling constant. Then, the entropy is transformed into $h(R_1')=h(R_1)+\ln|a|$, the variance becomes $\sigma_{1'}^2=a^2\sigma_1^2$, the covariances $\mathbf{\Gamma}_{1'j}=a \mathbf{\Gamma}_{1j}$, and the commutators $\left[R_1',R_j\right]=a\left[R_1,R_j\right]$. This implies that both $\mathbf \Gamma$ and $\mathbf C$ have one column and one row multiplied by $a$, so that  $\det\mathbf\Gamma$ and $\det\mathbf C$ are both multiplied by $a^2$. Inserting these new values in Eq.~(\ref{conjGenEUR}), we see that the constant term $\ln|a|$ appears on both sides of the inequality, confirming the invariance of this entropic uncertainty relation. 

It is straightforward to prove the validity of Eq.~(\ref{conjGenEUR}) for Gaussian states. Inserting the entropy of Gaussian-distributed variable $h(R_i) = \ln(2\pi e \sigma_i^2)/2$ into Eq.~(\ref{conjGenEUR}), we obtain
\eqarray{
	\frac{1}{2}\ln\left((2\pi e)^m\det\mathbf\Gamma\right)\geq\frac{1}{2}\ln((2\pi e)^m\det \mathbf C)
	\quad \Leftrightarrow\quad\det\mathbf\Gamma\geq\det \mathbf C
}
which is nothing else but the Robertson uncertainty relation, Eq.~(\ref{robGen}). 

The difficult (unresolved) problem is to prove  this conjecture for any state, not necessarily Gaussian. Here, we restrict ourselves to show that, for any state, Eq.~(\ref{conjGenEUR}) implies the Robertson uncertainty relation (\ref{robGen}) in its general form. The method works as usual. First, we use the entropy power of each variable $R_i$
\eq{N_i=\frac{1}{2\pi e}e^{2h(R_i)}\label{Ni}}
to rewrite  Eq. (\ref{conjGenEUR}) into its entropy-power form 
\eq{N_1N_2\cdots N_m  \, \frac{\det\mathbf\Gamma}  {\sigma_1^2\sigma_2^2\cdots \sigma_m^2 } \geq \det \mathbf C\label{GenEURpuissanceEntropic}}
Then, we use the fact that, for a fixed variance, the maximum  entropy is given by a Gaussian distribution, that is,  $N_i\leq\sigma_i^2$. We thus obtain the chain of inequalities
\eq{ \det\mathbf\Gamma    \geq   N_1N_2\cdots N_m \, \frac{\det\mathbf\Gamma} {\sigma_1^2\sigma_2^2\cdots \sigma_m^2}  \geq \det\mathbf C}
from which we deduce Eq. (\ref{robGen}).

Now, in order to avoid the problem that the entropic uncertainty relation (\ref{conjGenEUR}) is only defined for $m\leq 2n$, we may relax the bound by ignoring the correlations between the $R_i$'s as characterized by $\mathbf \Gamma$. This leads to the following (weaker) relation:
\begin{conj}   \label{conj2}
	Any $n$-modal state $\rho$ satisfies the entropic uncertainty relation
\eq{h(R_1)+h(R_2)+\cdots+h(R_m)\geq\frac{1}{2}\ln((2\pi e)^m\det\mathbf C)\label{relaxedConj}}	
	where the $R_i$'s are $m$ arbitrary continuous observables and $\mathbf C$ is the matrix of commutators as  defined  in Eq. (\ref{sigmaC}).
\end{conj}
This entropic uncertainty relation may probably be easier to prove than Eq.~(\ref{conjGenEUR}). From its entropy-power formulation, we see that it implies the weaker form of the Robertson uncertainty relation, Eq.~(\ref{robweak}), that is, we have the chain of inequalities
\eq{\sigma_1^2\sigma_2^2\cdots \sigma_m^2 \geq N_1N_2\cdots N_m\geq \det \mathbf C.\label{lessGenEURpuissanceEntropic}}
It is also immediate to see that Eq.~(\ref{conjGenEUR}) implies Eq.~(\ref{relaxedConj}) as a result of Hadamard inequality, Eq.  (\ref{hadam}), so that Conjecture \ref{conj2} is indeed weaker than Conjecture \ref{conj1}.

A problem with these two conjectures remains that they are irrelevant for an odd number $m$ of observables. We then conjecture a third version of an entropic uncertainty relation which holds for any $m$, but only for one-mode states ($n=1$):

\begin{conj}
\label{conj3}
Let $\mathbf{R}=(\hat R_1,\cdots,\hat R_m)$ be a vector of $m$ continuous observables acting on one mode as $\mathbf R=\mathbf a \hat x+\mathbf b \hat p$,
with $\hat x$ and $\hat p$ being the canonically conjugate quadratures of the mode as in Eq. (\ref{defab}).
Then, any one-mode state $\rho$ satisfies the entropic uncertainty relation
\begin{equation}
h(R_1)+h(R_2)+\cdots+h(R_m)\geq\frac{m}{2}\ln\left(\frac{2\pi e  }{m}|\mathbf a\wedge \mathbf b|\right)
\label{conjStefan}
\end{equation}
where the norm of the wedge product between vectors $\mathbf a$ and $\mathbf b$ is computed with Eq.~(\ref{wedgeproduct}).
\end{conj}
Its entropy-power form is
\eq{N_1N_2\cdots N_m\geq\left(\frac{|\mathbf a\wedge\mathbf b|}{m}\right)^m\label{EFstefan}}
where the $N_i$ are defined as in Eq. (\ref{Ni}). From Eq. (\ref{EFstefan}), we can deduce the variance-based uncertainty relation Eq.~(\ref{eqStefan}), which was derived in \cite{Weigert2}.

Let us mention that, for $m=2$, Conjecture \ref{conj3} reduces to the entropic uncertainty relation (\ref{huang}) for two arbitrary quadratures in the special case of one mode ($n=1$), so it is proven \cite{huang}. Indeed, if $\hat R_1=a_1 \hat x +b_1 \hat p$ and $\hat R_2=a_2 \hat x +b_2 \hat p$, we have
\begin{equation}
|\mathbf a\wedge\mathbf b| = | a_1 b_2 - a_2 b_1 | = | [ \hat R_1, \hat R_2] | \, ,
\end{equation}
so that Eq. (\ref{conjStefan}) becomes identical to Eq. (\ref{huang}). Furthermore, Conjecture \ref{conj3} reduces to the entropic uncertainty relation (\ref{eq-Guanlei}) for two rotated quadratures, that is, when we choose $\mathbf{a}=(\cos\theta,\sin\theta)$ and $\mathbf{b}=(\cos\phi,\sin\phi)$, so that $|\mathbf a\wedge\mathbf b|=|\sin(\theta-\phi)|$.

Finally, let us consider the special case of Eq. (\ref{defRpolygon}), where we have $m$ quadratures that are equidistributed around the unit circle, that is 
\eq{\hat R_i=\cos\phi_i\, \hat x+\sin \phi_i \,\hat p\qquad\text{with}\quad\phi_i=\frac{2\pi(i-1)}{m},\qquad i=1,\dots,m. \label{defxipolygon}  }
In this case, Conjecture \ref{conj3} reduces to the following entropic uncertainty relation, which was already conjectured in \cite{Weigert2}, and which we prove here\footnote{The proof is conditional on one reasonable assumption, see below}, namely
\begin{conj}   \label{conj4}
Let $\mathbf{R}=(R_1,\cdots,R_m)$ be a vector of $m$ continuous observables acting on one mode and equidistributed as defined in Eq.~(\ref{defxipolygon}).
Then, any one-modal state $\rho$ satisfies the entropic uncertainty relation \cite{Weigert2}
\eq{h(R_1)+h(R_2)+\cdots+h(R_m)\geq\frac{m}{2}\ln\left(\pi e \right).
	\label{conjStefanPolygon}}
\end{conj}
Indeed, for equidistributed $R_i$'s, we have $|\mathbf a\wedge\mathbf b|=m/2 $ as shown in Eq. (\ref{eq-wedge-commutators}), so that  Eq. (\ref{conjStefan})
reduces to Eq.~(\ref{conjStefanPolygon}). Similarly as before, its entropy-power form is
\eq{N_1N_2\cdots N_m\geq\left(\frac{1}{2}\right)^m  ,  \label{EFstefanpoly}}
where the $N_i$ are defined as in Eq.(\ref{Ni}) and, from it, we can deduce the variance-based uncertainty relation (\ref{eqstefanpoly}) as derived in \cite{Weigert2}.

We present here a partial proof of Conjecture \ref{conj4} using a variational method, following the same lines as in Section \ref{proofEURgen}, which is based on the extremization of our functional
\eq{F(\rho)=h(R_1)+\cdots+h(R_m).}
After proving that the vacuum state is a local extremum of $F(\rho)$, we will again assume that it is its global minimum. 
This assumption seems reasonable since the vacuum minimizes the corresponding variance-based uncertainty relation (\ref{eqstefanpoly}) as shown in \cite{Weigert2}.
Let us consider pure states $\rho=\ket{\psi}\bra{\psi}$ first. Here too, our functional $F(\rho)$ is invariant under displacements, so we can restrict to states centered on the origin. Inserting the constraints of normalization and zero mean values, we want to solve $\frac{\partial J}{\partial\bra{\psi}}=0$, where
\eq{J=h(R_1)+\cdots+h(R_m)+\lambda(\mean{\psi|\psi}-1)+\sum_{i=1}^m\mu_i\mean{\psi|\hat R_i|\psi_i}}
and $\lambda$ and $\mu_i$ are Lagrange multipliers. As shown in Section \ref{proofEURgen},
\eqarray{&& \frac{\partial h(R_i)}{\partial \bra{\psi}} = -\Big( \ln P(R_i)+1 \Big) \ket{\psi}\nonumber\\
&&  \frac{\partial}{\partial\bra{\psi}}\left(\lambda(\mean{\psi|\psi}-1)+\sum_{i=1}^m\mu_i\mean{\psi|\hat R_i |\psi}\right) = \left(\lambda+\sum_{i=1}^m\mu_i \hat R_i\right)\ket{\psi}}
so that the variational equation becomes
\eq{\left[-\ln \Big( P(R_1)P(R_2)\cdots P(R_m) \Big) -m+\lambda+\sum_{i=1}^m\mu_i \hat R_i\right]\ket{\psi}=0.  \label{new-eigenstate-equation} }
Thus, the eigenstates of Eq. (\ref{new-eigenstate-equation}) are the states extremizing the uncertainty functional. As before, instead of looking for eigenstates, we check that the vacuum state $\ket{0}$ is a solution of Eq. (\ref{new-eigenstate-equation}). It means that $P(R_i)$ is a Gaussian distribution
\eq{P(R_i)=\frac{1}{\sqrt{\pi}}e^{-R_i^2}}
where we used Eq.~(\ref{GaussDistN}) and the fact that the variance of $\hat R_i$ in the vacuum state is $1/2$ for all $\hat R_i$'s of Eq. (\ref{defxipolygon}). The variational equation can now be written as
\eq{\left[\frac{m}{2}\ln(\pi)+ \sum_{i=1}^m \hat R_i^2
	-m+\lambda+\sum_{i=1}^m\mu_i \hat R_i\right]\ket{0}=0}
which can be further simplified as
\eq{\left[\frac{m}{2}\ln(\pi)
	-\frac{m}{2}+\lambda+\sum_{i=1}^m\mu_i \hat R_i\right]\ket{0}=0   \label{furthersimplification}   }
by using the fact that
\eqarray{\sum_{i=1}^m \hat R_i^2\ket{\psi}&=&\sum_{i=1}^m\left[\cos\left(\frac{2\pi(i-1)}{m}\right)\hat x+\sin\left(\frac{2\pi(i-1)}{m}\right)\hat p\right]^2\ket{0}\nonumber\\
&=&\frac{m}{2}(\hat x^2+\hat p^2)\ket{0}\nonumber\\
&=&\frac{m}{2}\ket{0}.}
The value of $\lambda=\frac{m}{2}\ln(e/\pi)$ is found by multiplying Eq. (\ref{furthersimplification}) on the left by $\bra{0}$ and using the normalization condition as well as the fact that all mean values $\bra{0}\hat R_i\ket{0}$ must vanish. We are then left with $\sum_{i=1}^m\mu_i \hat R_i\ket{0}=0$, which is satisfied if $\mu_i=0$ for all $i$.
Thus, we have shown that there exists an appropriate choice of $\lambda$ and $\mu_i$ so that the vacuum extremizes  the uncertainty functional. Assuming that it is the global minimizer, we have proved Eq.~(\ref{conjStefanPolygon}) for pure states since $F(\ket{0})=\frac{m}{2}\ln(\pi e)$. Due to the concavity of the differential entropy, the entropic uncertainty relation (\ref{conjStefanPolygon}) then holds for mixed states too.

%
%
%
%
%
%


\section{Conclusion }

We have reviewed continuous-variable entropic uncertainty relations starting from the very first formulation by Hirschman and the proof by Białynicki-Birula and Mycielski to the recent entropic uncertainty relation between non-canonically conjugate variables, whose lower bound depends on the determinant of a matrix of commutators. We then showed that, by taking correlations into account, it is possible to define an entropic uncertainty relation for any two vectors of intercommuting quadratures whose minimum-uncertainty states are all pure Gaussian states. Finally, we derived several conjectures for an entropic uncertainty relation addressing more than $2$ continuous observables and gave a partial proof of one of them.

In Table \ref{implication}, we provide a summary of all entropic uncertainty relations (for Shannon differential entropies) encountered in this paper. These entropic uncertainty relations appear in the first column of Table \ref{implication}.  The symbol $\checkmark$ means that the relation is proven, $\dag$ that it is  proven conditionally on reasonable assumptions, and * that it is still a conjecture. As emphasized throughout this paper, entropic uncertainty relations are conveniently formulated in terms of entropy powers. The corresponding entropy-power uncertainty relations are then shown in the second column of Table \ref{implication}. Further, using the fact that the maximum entropy for a fixed variance is reached by a Gaussian distribution, a variance-based uncertainty relation can easily be deduced from each entropy-power uncertainty relation. This is what is done in the third column of Table  \ref{implication}, where we show the variance-based uncertainty relations that are implied by all entropy-power uncertainty relations.

We conclude this paper by noting that, although significant progress on entropic uncertainty relations has been achieved lately, we still lack a symplectic-invariant entropic uncertainty relation. All relations we have discussed are invariant under displacements (corresponding to a translation of the variables in phase space), and most of them are also invariant under squeezing transformations (corresponding to a scaling of the variables). However, no entropic uncertainty relation is invariant under rotations, which would make it invariant under the complete set of symplectic transformations. This is, however, a natural property of many variance-based uncertainty relations, such as the Robertson-Schrödinger relation~(\ref{robschr}). The invariance of the determinant of $\gamma$ makes the latter relation invariant under all symplectic transformations, hence under all Gaussian unitaries (since it is invariant under displacements too). A mentioned in Section \ref{EURxpcorrelations},
the joint entropy $h(x,p)$ would have the desired property to build a symplectic-invariant entropic uncertainty relation, but it is not defined for states with a negative Wigner function (see also \cite{hertz2,these}). The tight entropic uncertainty relations of Eqs. (\ref{nmodes}) and (\ref{NEWEUR}) admit all pure Gaussian states as minimum-uncertainty state, but are nevertheless not invariant under rotations.
A recent attempt at defining a symplectic-invariant entropic uncertainty relation is made in \cite{hertz4}, which builds on a multi-copy uncertainty observable that is related to the Schwinger representation of a spin state via harmonic oscillators.


\renewcommand\arraystretch{1.7}
\begin{sidewaystable}[h!]
	\begin{center}
		\begin{tabular}{c|c|c}
			\hline
			\hline
			Entropic UR & Entropy-power UR &Variance-based UR\\
			\hline
			\hline
			$h(x)+h(p)\geq \ln (\pi e)$&   $	N_x \, N_p \ge \frac{1}{4}  \,  $  &$\sigma_x^2\sigma_p^2\geq\frac{1}{4}$\\
			\hspace{15pt} Białynicki-Birula and Mycielski for $n=1$, Eq. (\ref{birulabis}) \hspace{0pt} $\checkmark$&  		Eq.	(\ref{EPUR})   &Heisenberg, Eq. (\ref{heis})\\
			\hline
			$h(\mathbf{x})+h(\mathbf{p})-\frac{1}{2}\left(\frac{\det \gamma_{\mathbf{x}}\det\gamma_{\mathbf{p}}}{\det \gamma}\right)\geq n\ln (\pi e)$&  $	N_{\mathbf{x}} N_{\mathbf{p}}
			\left( \frac{\det\gamma}{\det\gamma_{\mathbf{x}}\,\det\gamma_{\mathbf{p}}} \right)^{1/n} \geq\frac{1}{4} \, $  &$\det\gamma\geq\frac{1}{4^{n}}$\\
			\hspace{113pt} Eq. (\ref{nmodes}) \hspace{113pt} $\dag$&  Eq. (		\ref{n-mode-conjecture-bis})   &
			Eq. (\ref{RSanmodes})\\
			\hline
			$h(\mathbf{y})+h(\mathbf{z})\geq \ln \left((\pi e)^n|\det \mathbf{K}|\right) $&  $N_{\mathbf{y}} N_{\mathbf{z}} \geq \frac{ |\det \mathbf{K}|^{2/n} }{4} $   &$\det \mathbf{\Gamma}_{\mathbf{y}} \,  \det \mathbf{\Gamma}_{\mathbf{z}}\geq\frac{|\det \mathbf{K}|^2}{4^n}$\\
			\hspace{118pt}Eq. (\ref{EURFRFTcom})\hspace{115pt} $\checkmark$&   Eq. (	\ref{entropy-power-ur})    &Eq. (\ref{URcovariance})\\
			\hline
			$	h({\bf y})+h({\bf z})-\frac{1}{2}\ln\left(\frac{\det\mathbf{\Gamma}_{\mathbf{y}}\det\mathbf{\Gamma}_{\mathbf{z}}}{\det\mathbf{\Gamma}}\right)\geq\ln\left((\pi e)^n | \det \mathbf{K}|\right)$&  $	N_{\mathbf{y}}N_{\mathbf{z}} \left( \frac{\det \mathbf{\Gamma}}{\det\mathbf{\Gamma}_{\mathbf{y}}\det\mathbf{\Gamma}_{\mathbf{z}}} \right)^{1/n} \geq \frac{ |\det \mathbf{K}|^{2/n} }{4}
			$ &$\det\mathbf{\Gamma}\geq\frac{ |\det \mathbf{K}| ^2}{4^n}$\\
			\hspace{118pt} Eq. (\ref{NEWEUR})\hspace{115pt} $\dag$&   Eq. (\ref{entropy-power-ur2})  & Eq. (\ref{RobGen})\\
			\hline
			\hline
			$h(R_1)+\cdots+h(R_m)-\frac{1}{2}\ln\left(\frac{\sigma_1^2\sigma_2^2\cdots\sigma_m^2}{\det\mathbf{\Gamma}}\right)\geq\frac{1}{2}\ln((2\pi e)^m\det \mathbf{C})$& $	N_1\cdots N_m\frac{\det\mathbf{\Gamma}}{  \sigma_1^2\sigma_2^2\cdots \sigma_m^2  }\geq \det \mathbf{C}$ &$\det\mathbf{\Gamma} \geq\det \mathbf{C}$ \\
			\hspace{110pt} Eq. (\ref{conjGenEUR}) \hspace{110pt} *& Eq. (\ref{GenEURpuissanceEntropic})&Robertson, Eq. (\ref{robGen})\\
			\hline
				$h(R_1)+\cdots+h(R_m)\geq\frac{1}{2}\ln((2\pi e)^m\det \mathbf{C})$& $	N_1\cdots N_m\geq \det \mathbf{C}$ &$\sigma_1^2\cdots\sigma_m^2 \geq\det \mathbf{C}$ \\
			\hspace{110pt} Eq. (\ref{relaxedConj}) \hspace{110pt} *& Eq. (\ref{lessGenEURpuissanceEntropic})&Robertson, Eq. (\ref{robweak})\\
			\hline
			$h(R_1)+\cdots+h(R_m)\geq\frac{m}{2}\ln\left(\frac{2\pi e  }{m}|\mathbf a\wedge \mathbf b|\right)$&$N_1\cdots N_m\geq\left(\frac{|\mathbf a\wedge\mathbf b|}{m}\right)^m$&$\sigma_1^2\cdots\sigma_m^2\geq\left(\frac{|\mathbf a\wedge\mathbf b|}{m}\right)^m$\\
			\hspace{110pt} Eq. (\ref{conjStefan})\hspace{110pt} *& Eq. (\ref{EFstefan})& Eq. (\ref{eqStefan})\\
			\hline
			$h(R_1)+\cdots+h(R_m)\geq\frac{m}{2}\ln\left(\pi e \right)$&$N_1\cdots N_m\geq \frac{1}{2^m}$&$\sigma_1^2\cdots\sigma_m^2\geq\frac{1}{2^m}$\\
		\hspace{110pt} Eq. (\ref{conjStefanPolygon})\hspace{110pt} $\dag$& Eq. (\ref{EFstefanpoly})& Eq. (\ref{eqstefanpoly})\\
		\hline
		\end{tabular}
		\caption[Summary of the entropic uncertainty relations]{\label{implication}Summary of the entropic uncertainty relations (UR) expressed in terms of differential entropies (first column), their corresponding entropy-power formulations (second column), as well as their implied variance-based uncertainty relations (third column). The symbol $\checkmark$ means that the entropic uncertainty relation is proven, $\dag$ that it is proven conditionally on reasonable assumptions and * that it is  a conjecture. We set $\hbar=1$ in all equations.}
	\end{center}
\end{sidewaystable}
\renewcommand\arraystretch{1}

\medskip
\section*{Acknowlegments}
We thank Michael Jabbour and Luc Vanbever for their collaboration and Stefan Weigert for helpful discussions.
This work was supported by the F.R.S.-FNRS Foundation under Projects No. T.0199.13 and T.0224.18.
A.H. acknowledges financial support from the F.R.S.-FNRS Foundation.


\appendix
\setcounter{section}{0}
\section{Symplectic formalism}
\label{appA}

Here, we briefly review the representation of Gaussian states and unitaries in phase space based on the symplectic formalism (see also \cite{weed}).
A $n$-mode Gaussian state $\rho$ has a Gaussian Wigner function of the form 
\eq{W_G(\mathbf{x,p})=\frac{1}{(2\pi)^n\sqrt{\det \gamma}}e^{-\frac{1}{2}(\mathbf{r-\mean{r}})^T\gamma^{-1}(\mathbf{r-\mean{r}})}.
	\label{wignergauss}}
 and is completely 
 characterized by its {\it vector of mean values} $\mean{\mathbf{r}}=\tr(\mathbf{r}\rho)$
 and its {\it covariance matrix} $\gamma$, whose elements are given by
 \eq{\gamma_{ij}=\frac{1}{2}\mean{\{\hat r_i,\hat r_j\}}-\mean{\hat r_i}\mean{\hat r_j}.\label{covmatdef}}
Here, ${\bf r}=(\hat x_1,\hat p_1,\hat x_2,\hat p_2,\cdots,\hat x_n,\hat p_n)$ is the quadrature vector, $\mean{\cdot}$ stands for the expectation value $\tr(\cdot\rho)$, and $\{\cdot,\cdot\}$ stands for the anti-commutator. Remark that the covariance matrix $\gamma$ is a real, symmetric, and positive semi-definite matrix. It must also comply with the uncertainty relation
 \eq{\gamma+i{\Omega\over 2}\geq 0   \label{equA3} }  
where $\Omega$ is the so-called symplectic form, defined as 
 \eq{ \Omega=\bigoplus\limits_{k=1}^n \omega,\qquad\omega=\begin{pmatrix}
 		0&1\\-1&0	\end{pmatrix} .
 	\label{omega}}
Equation~(\ref{equA3}) is a necessary and sufficient condition that $\gamma$ has to fulfill in order to be the covariance matrix of a physical  state \cite{simon94}.
 
The  purity $\mu$ of a Gaussian state is given by
\eq{\mu_G=\frac{1}{2^n\sqrt{\det \gamma}}}
and it can be shown that pure states having $\det\gamma=1/4^n$ are necessarily Gaussian.

The  simplest example of a one-modal Gaussian state is the {\it vacuum} state $\ket{0}$. It has a vector of mean values equal to $(0,0)^T$ and its covariance matrix is given by $\gamma_{vac}=\mathds{1}/2$.
We can displace the vacuum in phase state by applying a Gaussian unitary resulting in another Gaussian state called a {\it coherent state} $\ket{\alpha}=D(\alpha)\ket{0}$, where $D(\alpha)=e^{\alpha \hat a^\dag-\alpha^*\hat a}$ and $\hat a$ is the annihilation operator. The covariance matrix of a coherent state is the same as for the vacuum, but its vector of mean values changes as $\mean{\mathbf{r}}_\alpha=\sqrt{2}\begin{psmallmatrix}
\Re ( \alpha)\\ \Im (\alpha)
\end{psmallmatrix}$.
As another Gaussian unitary, we can squeeze the variance of a quadrature and obtain another Gaussian state known as a {\it squeezed state} $\ket{z}=S(z)\ket{0}$ with $S(z)=e^{\frac{1}{2}(z^*\hat a^2-z\hat a^{\dag2})} $, where $z=re^{i\phi}$ is a complex number ($r$ is the squeezing parameter and $\phi$ the squeezing angle). The symplectic matrix associated to this Gaussian unitary is given by 
\eq{\mathcal{S}_z=\left(
	\begin{array}{cc}
		\cosh r-\cos 2 \phi  \sinh r& -\sin 2 \phi  \sinh r \\
		-\sin 2 \phi  \sinh r & \cosh r+\cos 2 \phi  \sinh r \\
	\end{array}
	\right)}
so that the covariance matrix of a squeezed state is given by \eq{\gamma_z=\mathcal{S}_z\gamma_{vac}\,\mathcal{S}_z^T= \frac{1}{2} \begin{pmatrix}
	\cosh 2r-\cos2\phi\,\sinh 2r&-\sin2\phi\,\sinh 2r\\-\sin2\phi\,\sinh 2r&\cosh 2r+\cos2\phi\,\sinh 2r
	\end{pmatrix}.}
A squeezing in the $x$ ($p$) direction corresponds to the choice $\phi=0$ ($\phi=\pi/2$).
Yet another (one-mode) Gaussian operation is the phase-shift  $R(\theta)=e^{-i\theta\hat a^\dag \hat a}$. Its associated symplectic matrix is simply given by the rotation matrix
\eq{\mathcal{R}_\theta=\begin{pmatrix}
		\cos\theta&\sin\theta\\-\sin\theta&\cos\theta
	\end{pmatrix}.}

The above operations are all the possible one-modal Gaussian unitaries. For $n$~modes, there is a larger set of Gaussian unitary operations, which we do not need to discuss here. The key point is that, in state space, a Gaussian unitary always transforms a Gaussian state onto a Gaussian state. Its corresponding action in phase space is expressed via a symplectic transformation.
That is, if a Gaussian unitary $U$ transforms $\rho$ according to
\eq{\hat \rho\rightarrow U\hat{\rho}\,U^\dag}
its quadratures in phase space are transformed as 
\eq{{\bf \hat r}\rightarrow \mathcal{S} \bf{\hat r+d}}
where $\bf d$ is a real vector of dimension $2n$ and $\mathcal{S}$ is a real $2n\times 2n$ matrix. Regarding the mean values and covariance matrix of $\rho$, the transformation rules are
\eq{{\bf \mean{r}}\rightarrow  \mathcal{S} \mathbf{\mean{r}+d}\qquad \text{and}\qquad\gamma\rightarrow  \mathcal{S}\gamma  \mathcal{S}^T.
}
The commutation relations between the quadratures have to be preserved along this transformation, which is the case if the matrix $\mathcal{S}$ is {\it symplectic}, that is, if
\eq{ \mathcal{S}\Omega  \mathcal{S}^T=\Omega}
where $\Omega$ is defined in Eq.~(\ref{omega}). Note that $\Omega^T=\Omega^{-1}=-\Omega$ and $\Omega^2 =-\mathds{1}$.
Be aware that this definition of symplectic matrices is linked to the definition of $\mathbf{r}$ (i.e., the ordering of the entries in $\mathbf{r}$). If one chooses instead to define ${\bf r}=(\hat x_1,\cdots,\hat x_n,\hat p_1,\cdots,\hat p_n)$, then the matrix $\mathcal{S}$ is symplectic if
$ \mathcal{S}J  \mathcal{S}^T=J$ with  $J=\begin{psmallmatrix}
		0&\mathds{1}\\-\mathds{1}&0
	\end{psmallmatrix}$.
Here too,  $J^T=J^{-1}=-J$ and $J^2 =-\mathds{1}$.

In addition, any symplectic matrix $\mathcal{S}$ has the following properties:
\begin{itemize}
	\item The matrices $\mathcal{S}^T$, $\mathcal{S}^{-1}$  and $-\mathcal{S}$ are also symplectic.
	\item The inverse of $\mathcal{S}$ is given by $\mathcal{S}^{-1}=-\Omega \mathcal{S}^T\Omega\,\,\,$ (or $\mathcal{S}^{-1}=-J \mathcal{S}^TJ$, depending on the  definition of $\mathbf{r}$).
	\item $\det\mathcal{S}=1$, which implies that $\det\gamma$ is conserved by any symplectic transformation.
	\item If ${\bf r}=(\hat x_1,\cdots,\hat x_n,\hat p_1,\cdots,\hat p_n)$  and $\mathcal{S}=\begin{psmallmatrix}
	a&b\\c&d
	\end{psmallmatrix}$, then $\mathcal{S}J\mathcal{S}^T=J$ implies that $ab^T$ and $cd^T$ are symmetric matrices and $ad^T-bc^T=\mathds{1}$.
	\item In terms of the associated symplectic transformation, a Gaussian unitary will be passive (it conserves the mean photon number) if and only if
	\eq{{\bf d}=0\qquad\text{and}\qquad  \mathcal{S}^T \mathcal{S}=\mathds{1},}
	which means that the symplectic matrix $\mathcal{S}$ must be orthogonal.
\end{itemize}  

The {\it Williamson's theorem} \cite{Williamson}  states that, after the appropriate symplectic transformation, every real, positive semidefinite matrix of even dimension can be brought to a diagonal form $\gamma^\oplus$, with its  {\it symplectic values} $\nu_k$ on the diagonal (each $\nu_k$ is doubly degenerate). In other words, there exists a symplectic matrix $\mathcal{S}$ such that\footnote{We use here the definition $\mathbf{r}=(\hat x_1,\hat p_1,\cdots,\hat x_n,\hat p_n)$. }
\eq{\gamma=\mathcal{S}\gamma^\oplus\mathcal{S}^T,\qquad\text{where}\qquad \gamma^\oplus=\bigoplus\limits_{k=1}^{n}\nu_k \, \mathds{1}_{2\times2}.\label{symplecticmatrix}}
Obviously, since the determinant of a symplectic matrix is equal to $1$, $\gamma$ and $\gamma^\oplus$ have the same determinant.
Therefore, for a one-mode state, its symplectic value is simply equal to $\sqrt{\det \gamma}$. For a two-mode state, the two symplectic values $\nu_\pm$ can be found using the following formula \cite{serafini}
\eq{\nu_\pm=\sqrt{\frac{\Delta\pm\sqrt{\Delta^2-4\det\gamma}}{2}}\label{symplectic2mode}}
where the covariance can be written in the block form
\eq{\gamma=\begin{pmatrix}
		A&C\\C^T&B
\end{pmatrix}}
and $\Delta=|A|+|B|+2|C|$.
In general, one can find the symplectic values by diagonalizing the matrix $i\Omega \gamma$ and taking the absolute value of its eigenvalues (see e.g. \cite{weed,these}).


\section{Calculation of $\text{det}\,\mathbf{\Gamma}$}
\label{appB}

Here, we compute the determinant of the covariance matrix $\mathbf{\Gamma}$ of the $y,z$-quadratures [see Eq. (\ref{gammayz})] as needed in the evaluation of the 
tight entropic uncertainty relation for two arbitrary vectors of quadratures, Eq. (\ref{NEWEUR}).
As before, let  $\mathbf y = (\hat y_1,\cdots \hat y_n)^T$  be a vector of commuting quadratures and $\mathbf z =~(\hat z_1,\cdots \hat z_n)^T$ be another vector of commuting quadratures. We now suppose that they correspond to the output $x$-quadratures after applying two symplectic transformations denoted as
$\mathcal{A}$ and $\mathcal{B}$ onto the $2n$-dimensional vector of input quadratures
$\mathbf{r}=(\hat x_1,\cdots ,\hat x_n,\hat p_1,\cdots ,\hat p_n)^T$. The corresponding $2n$-dimensional vectors of output quadratures are written as
 \begin{equation}
	\mathbf{r}_A =\mathcal{A} \; \mathbf{r} 
	\equiv \begin{pmatrix} \mathbf y \\ \mathbf q \end{pmatrix},
	\qquad
	\mathbf{r}_B =\mathcal{B} \; \mathbf{r}
	\equiv \begin{pmatrix} \mathbf z \\ \mathbf o \end{pmatrix} 
	\label{defyz2}
\end{equation}
where $\mathbf q$ (resp. $\mathbf o$) is the vector of quadratures that are canonically conjugate with $\mathbf y$ (resp. $\mathbf z$). 
Since Eq.~(\ref{defyz2}) tells us how to obtain $\mathbf y$ and $\mathbf z$ from $\mathbf x$ and $\mathbf p$ through the symplectic transformations $\mathcal{A}$ and $\mathcal{B}$ , we can compute the  elements of the covariance matrix $\mathbf{\Gamma}$ for the  $y,z$ quadratures, namely
\begin{equation}
	\mathbf{\Gamma}_{ij}={1\over 2}\mean{\hat R_i\hat R_j+\hat R_j\hat R_i}-\mean{\hat R_i}\mean{\hat R_j}
\end{equation}
with ${\bf R}=(\hat y_1,...,\hat y_n,\hat z_1,...,\hat z_n)^T$. For example, we may evaluate $\mathbf{\Gamma}_{ij}$ for $1\leq i,j\leq n$, namely
\eqarray{\mathbf{\Gamma}_{ij}&=&{1\over 2}\mean{\sum_{k=1}^{2n}\mathcal{A}_{ik}\hat r_k\,\sum_{l=1}^{2n}\mathcal{A}_{jl}\hat r_l+\sum_{l=1}^{2n}\mathcal{A}_{jl}\hat r_l\,\sum_{k=1}^{2n}\mathcal{A}_{ik}\hat r_k}-\mean{\sum_{k=1}^{2n}\mathcal{A}_{ik}\hat r_k}\mean{\sum_{l=1}^{2n}\mathcal{A}_{jl}\hat r_l}\nonumber\\
	&=&	\sum_{k=1}^{2n}	\sum_{l=1}^{2n}\mathcal{A}_{ik}\mathcal{A}_{jl}\gamma_{kl}\nonumber\\
	&=&	\sum_{k=1}^{2n}	\sum_{l=1}^{2n}\mathcal{A}_{ik}\gamma_{kl}\mathcal{A}^T_{lj}\nonumber\\
	&=&(\mathcal{A}\gamma\mathcal{A}^T)_{ij}.
}
Similarly, we can show that  $\mathbf{\Gamma}_{i+n,j+n}=(\mathcal{B}\gamma\mathcal{B}^T)_{ij}$ and $\mathbf{\Gamma}_{i,j+n}=(\mathcal{A}\gamma\mathcal{B}^T)_{ij}$ for $1\leq i,j\leq n$.
Since the covariance matrix is symmetric, we obtain
\begin{equation}
\mathbf{\Gamma}=\begin{pmatrix}
\mathbf{\Gamma}_{\mathbf{y}}&\mathbf{\Gamma}_{yz}\\\mathbf{\Gamma}_{yz}&\mathbf{\Gamma}_{\mathbf{z}}
\end{pmatrix}=\begin{pmatrix}
(\mathcal{A}\gamma\mathcal{A}^T)_{i,j=1,...,n}&(\mathcal{A}\gamma\mathcal{B}^T)_{i,j=1,...,n}\\(\mathcal{B}\gamma\mathcal{A}^T)_{i,j=1,...,n}&(\mathcal{B}\gamma\mathcal{B}^T)_{i,j=1,...,n}
\end{pmatrix}.
\end{equation}
Notice that matrices $\mathcal{A}\gamma\mathcal{A}^T$, $\mathcal{A}\gamma\mathcal{B}^T$, $\mathcal{B}\gamma\mathcal{A}^T$ and $\mathcal{B}\gamma\mathcal{B}^T$ all have dimensions $2n\times 2n$ but we truncate them to keep only the reduced matrices with indices running from $1$ to $n$. Therefore, $\mathbf{\Gamma}_{\mathbf{y}}$, $\mathbf{\Gamma}_{\mathbf{z}}$ and $\mathbf{\Gamma}_{yz}$ have dimension $n\times n$, while $\mathbf{\Gamma}$ is a $2n\times 2n$ matrix.

To simplify the expression of $\mathbf{\Gamma}$, we use a block matrix representation of the symplectic transformations, 
\begin{equation}
\mathcal{A}=
\begin{pmatrix}
\mathcal{A}_a&\mathcal{A}_b\\\mathcal{A}_c&\mathcal{A}_d
\end{pmatrix}
\quad\text{and}\quad
\mathcal{B}=\begin{pmatrix}
\mathcal{B}_a&\mathcal{B}_b\\\mathcal{B}_c&\mathcal{B}_d
\end{pmatrix}
\label{matAB2}
\end{equation} 
so that, for example,
\begin{eqnarray}
\mathcal{A}\gamma\mathcal{A}^T&=&\begin{pmatrix}
\mathcal{A}_a&\mathcal{A}_b\\\mathcal{A}_c&\mathcal{A}_d
\end{pmatrix}\begin{pmatrix}
\gamma_{\mathbf{x}}&\gamma_{\mathbf{xp}}\\\gamma_{\mathbf{xp}}&\gamma_{\mathbf{p}}
\end{pmatrix}\begin{pmatrix}
\mathcal{A}_a^T&\mathcal{A}_c^T\\\mathcal{A}_b^T&\mathcal{A}_d^T
\end{pmatrix}\\
&=&\begin{pmatrix}
\mathcal{A}_a\gamma_{\mathbf{x}}\mathcal{A}_a^T+\mathcal{A}_a\gamma_{\mathbf{xp}}\mathcal{A}_b^T+\mathcal{A}_b\gamma_{\mathbf{xp}}\mathcal{A}_a^T+\mathcal{A}_b\gamma_{\mathbf{p}}\mathcal{A}_b^T&\cdots\\\cdots&\cdots
\end{pmatrix}\quad\nonumber
\end{eqnarray}
where we do not need to express the matrix elements denoted with dots since all we need to compute is
\begin{equation}
(\mathcal{A}\gamma\mathcal{A}^T)_{i,j=1,...,n}=\mathcal{A}_a\gamma_{\mathbf{x}}\mathcal{A}_a^T+\mathcal{A}_a\gamma_{\mathbf{xp}}\mathcal{A}_b^T+\mathcal{A}_b\gamma_{\mathbf{xp}}\mathcal{A}_a^T+\mathcal{A}_b\gamma_{\mathbf{p}}\mathcal{A}_b^T.
\end{equation}
By doing the same calculation for the other blocs of matrix $\mathbf{\Gamma}$, we obtain that it can be written as the product of three matrices,
\begin{equation}
\mathbf{\Gamma}=\begin{pmatrix}
\mathcal{A}_a&\mathcal{A}_b\\\mathcal{B}_a&\mathcal{B}_b
\end{pmatrix}\begin{pmatrix}
\gamma_{\mathbf{x}}&\gamma_{\mathbf{xp}}\\\gamma_{\mathbf{xp}}&\gamma_{\mathbf{p}}
\end{pmatrix}\begin{pmatrix}
\mathcal{A}_a&\mathcal{A}_b\\\mathcal{B}_a&\mathcal{B}_b
\end{pmatrix}^T.
\label{Gammagamma}
\end{equation}
In particular, the determinant of $\mathbf{\Gamma}$ is given by
\begin{eqnarray}
\det \mathbf{\Gamma}=\det \gamma\det\Bigg[\begin{pmatrix}
\mathcal{A}_a&\mathcal{A}_b\\\mathcal{B}_a&\mathcal{B}_b
\end{pmatrix}\Bigg]^2.
\end{eqnarray}
Note that for a block matrix $\mathbf{M}$ of size $(n+m)\times (n+m)$ written as
\eq{\mathbf{M}=\begin{pmatrix}
		\mathbf{A}_{n\times n}&\mathbf{B}_{n\times m}\\\mathbf{C}_{m\times n}&\mathbf{D}_{m\times m}
	\end{pmatrix},}
it is easy to see that the following equality holds (assuming that $\mathbf{D}$ is invertible\footnote{If $\mathbf{D}$ is not invertible, Eq. (\ref{decompo}) can be written in a similar way in terms of $\mathbf{A}^{-1}$. })
\eq{
	\begin{pmatrix} \mathbf{A}&\mathbf{B}\\\mathbf{C}&\mathbf{D}	\end{pmatrix}
	\begin{pmatrix} \mathds{1}&0\\-\mathbf{D}^{-1}\mathbf{C}&\mathds{1}	\end{pmatrix}=
	\begin{pmatrix} \mathbf{A}-\mathbf{B}\mathbf{D}^{-1}\mathbf{C}&\mathbf{B}\\0&\mathbf{D}	\end{pmatrix}.
	\label{decompo}
}
Thus, the determinant of this equation is 
\eq{\det(\mathbf{M})=\det(\mathbf{A}-\mathbf{B}\mathbf{D}^{-1}\mathbf{C})\det(\mathbf{D})\label{detM}}
where we have exploited the fact that the determinant of a block triangular matrix is given by
the product of the determinants of its diagonal blocks \cite{Silvester}.

Moreover, since $\mathcal{B}$ represents a symplectic transformation, it hence satisfies  $\mathcal{B}\begin{psmallmatrix} 0&\mathds{1}\\-\mathds{1}&0\end{psmallmatrix}\mathcal{B}^T=\begin{psmallmatrix} 0&\mathds{1}\\-\mathds{1}&0\end{psmallmatrix}$. In particular, this means that\footnote{$(\cdot)^{-T}$ denotes the transpose of the inverse.} $\mathcal{B}_a\mathcal{B}_b^T=\mathcal{B}_b\mathcal{B}_a^T$ or $\mathcal{B}_a=\mathcal{B}_b\mathcal{B}_a^T\mathcal{B}_b^{-T}$. Thus, using Eq.~(\ref{detM})  together with the symmetry of the matrix $\mathcal{B}_a\mathcal{B}_b^T$, we can  compute the determinant of the bloc matrix $\mathbf{M}$ in our case
\begin{eqnarray}
\det\Bigg[\begin{pmatrix}
\mathcal{A}_a&\mathcal{A}_b\\\mathcal{B}_a&\mathcal{B}_b
\end{pmatrix}\Bigg]&=&\det(\mathcal{A}_a-\mathcal{A}_b\mathcal{B}_b^{-1}\mathcal{B}_a)\det\mathcal{B}_b\nonumber\\
&=&\det(\mathcal{A}_a-\mathcal{A}_b\mathcal{B}_b^{-1}\mathcal{B}_b\mathcal{B}_a^T\mathcal{B}_b^{-T})\det \mathcal{B}_b^T\nonumber\\
&=&\det(\mathcal{A}_a\mathcal{B}_b^T-\mathcal{A}_b\mathcal{B}_a^T)\nonumber\\
&=&\det(\mathcal{B}_b\mathcal{A}_a^T-\mathcal{B}_a\mathcal{A}_b^T).
\end{eqnarray}
Thus, the determinant of $\mathbf{\Gamma}$ can be written in terms of the blocks composing the two symplectic transformations $\mathcal{A}$ and $\mathcal{B}$
\begin{equation}
\det\mathbf{\Gamma}=\det \gamma\left(\det(\mathcal{B}_b\mathcal{A}_a^T-\mathcal{B}_a\mathcal{A}_b^T)\right)^2.
\label{equivgammasymplectic}
\end{equation}

Now, this expression can be rewritten in a form that does not explicitly include the  blocks composing $\mathcal{A}$ and $\mathcal{B}$
but uses the commutator matrix $\mathbf{K}$ instead. The elements of $\mathbf{K}$ are expressed as
\begin{eqnarray}
\mathbf{K}_{ji}&=&[\hat y_j,\hat z_i]\nonumber\\
&=&\sum_{k=1}^{2n}\sum_{m=1}^{2n}\mathcal{A}_{jk}\mathcal{B}_{im}[\hat r_k,\hat r_m]\nonumber\\
&=&i\sum_{m=1}^{2n}\left(\sum_{k=1}^{n}\mathcal{A}_{jk}\mathcal{B}_{im}\delta_{m,k+n}-\sum_{k=n+1}^{2n}\mathcal{A}_{jk}\mathcal{B}_{im}\delta_{m,k-n}\right)\nonumber\\
&=&i\left(\sum_{k=1}^{n}\mathcal{A}_{jk}\mathcal{B}_{i,k+n}-\sum_{k=n+1}^{2n}\mathcal{A}_{jk}\mathcal{B}_{i,k-n}\right)\nonumber\\
&=&i\left(\sum_{k=1}^{n}\mathcal{A}_{jk}\mathcal{B}_{i,k+n}-\mathcal{A}_{j,k+n}\mathcal{B}_{ik}\right)\nonumber\\
&=&i\left(\sum_{k=1}^{n}(\mathcal{A}_a)_{jk}(\mathcal{B}_b)_{ik}-(\mathcal{A}_b)_{jk}(\mathcal{B}_a)_{ik}\right)\nonumber\\
&=& i\left(\mathcal{B}_b\mathcal{A}_a^T-\mathcal{B}_a\mathcal{A}_b^T\right)_{ij},
\end{eqnarray}
which implies that
\eq{ |\det \mathbf{K}| = |\det(\mathcal{B}_b\mathcal{A}_a^T-\mathcal{B}_a\mathcal{A}_b^T)| } 
as proven in \cite{hertz3}. 
Hence, Eq. (\ref{equivgammasymplectic}) can finally be expressed as
\begin{equation}
\det \mathbf{\Gamma}=\det \gamma| \det \mathbf{K}|^2.
\label{equivgammaKapp}
\end{equation} 
that is, Eq. (\ref{equivgammaK}). 


\section{Pure Gaussian states as eigenvectors of ${1\over 2} \,{\bf R} ^T\mathbf{\Gamma}^{-1}{\bf R}$}
\label{appC}

Here, we show that $n$-modal pure Gaussian states are eigenvectors of the operator ${1\over 2} \,{\bf R} ^T\mathbf{\Gamma}^{-1}{\bf R}$ with eigenvalue $n$, see Eq.~(\ref{eigenequ3bis}).
In state space, a pure Gaussian state  can be written as $\ket{\psi^G}=\hat S\ket{0}$,  where $\hat S$
is a Gaussian unitary and $\ket{0}$ is the $n$-modal vacuum state. Since the states considered in the proof of Eq. (\ref{NEWEUR}) are centered at the origin, we do not need to apply a displacement operator and $\hat S$ is a $n$-modal squeezing operator (with arbitrary squeezing and rotation). 
In order to apply  ${1\over 2} \,{\bf R} ^T\mathbf{\Gamma}^{-1}{\bf R}$ onto state $\ket{\psi^G}$, we write the canonical transformation of ${\bf r}$ in phase space that corresponds to $\hat S$ in state space (in the Heisenberg picture), namely
${\hat S}^{\dagger} {\bf r}\,{\hat S} =\mathcal M {\bf r}$, where $\mathcal M$ is a symplectic matrix
so that  $\gamma^G = \mathcal M \gamma_\mathrm{vac} \mathcal M^T$. Remember that $\gamma^G$ is the covariance matrix for the $x,p$-quadratures, but we are interested in the covariance matrix $\mathbf{\Gamma}$ for the $y,z$ quadratures. We thus use the following change of variables
\begin{equation}
\hat S^\dag \begin{pmatrix}{\bf y}\\ {\bf z}\end{pmatrix}\hat S =
\hat S^\dag \begin{pmatrix}
\mathcal{A}_a&\mathcal{A}_b\\\mathcal{B}_a&\mathcal{B}_b
\end{pmatrix} \begin{pmatrix}{\bf x}\\{\bf p}\end{pmatrix}\hat S =
\begin{pmatrix}
\mathcal{A}_a&\mathcal{A}_b\\\mathcal{B}_a&\mathcal{B}_b
\end{pmatrix} \hat S^\dag \begin{pmatrix}{\bf x}\\{\bf p}\end{pmatrix}\hat S =
\begin{pmatrix}
\mathcal{A}_a&\mathcal{A}_b\\\mathcal{B}_a&\mathcal{B}_b
\end{pmatrix}\mathcal M \begin{pmatrix}{\bf x}\\{\bf p} \end{pmatrix}
\end{equation}
where we have use the fact that $\hat S$ and $\begin{psmallmatrix}
\mathcal{A}_a&\mathcal{A}_b\\\mathcal{B}_a&\mathcal{B}_b
\end{psmallmatrix}$ commute since they act on two different spaces. Then, we have
\begin{eqnarray}
\hspace{-1cm}  {1\over 2} {\bf R} ^T\mathbf{\Gamma}^{-1} \, \,{\bf R} \, \,\,\ket{\psi^G}
&=&{1\over 2} \, \,{\bf R}^T\, \mathbf{\Gamma}^{-1} \,  \,{\bf R}\, \hat S \ket{0}\nonumber\\
&=&
\frac{1}{2}{\hat S} \hat S^\dag\, \,{\bf R}^T \hat S\mathbf{\Gamma}^{-1} \hat S^\dag\,{\bf R}\hat S\,\ket{0}\nonumber\\
&=&
\frac{1}{2}{\hat S} \begin{pmatrix}\hat S^\dag {\bf y}\hat S&\hat S^\dag {\bf z}\hat S \end{pmatrix}
\mathbf{\Gamma}^{-1} \,\begin{pmatrix}\hat S^\dag {\bf y}\hat S\\\hat S^\dag {\bf z}\hat S \end{pmatrix}\ket{0}\nonumber\\
&=&
\frac{1}{2}{\hat S}\begin{pmatrix}{\bf x}&{\bf p} \end{pmatrix}\mathcal M^T\begin{pmatrix}
\mathcal{A}_a&\mathcal{A}_b\\\mathcal{B}_a&\mathcal{B}_b
\end{pmatrix}^T\mathbf{\Gamma}^{-1}
\begin{pmatrix}
\mathcal{A}_a&\mathcal{A}_b\\\mathcal{B}_a&\mathcal{B}_b
\end{pmatrix}
\mathcal M\begin{pmatrix}{\bf x}\\{\bf p} \end{pmatrix} \,  \ket{0} \nonumber\\
&=&\frac{1}{2}{\hat S}   \,{\bf r}^T\,\gamma_\mathrm{vac}^{-1} \,{\bf r} \,  \ket{0}  \nonumber\\
&=&{\hat S}   \, (|{\bf x}|^2+|{\bf p}|^2) \,  \ket{0}  \nonumber\\
&=&n\, {\hat S}\ket{0} = n\, \ket{\psi^G}.
\label{eigen}
\end{eqnarray}
To find the fifth line, we have used Eq.~(\ref{Gammagamma}) in order to compute the inverse of $\mathbf{\Gamma}$, namely
\begin{equation}
\mathbf{\Gamma}^{-1}= \left[\begin{pmatrix}
\mathcal{A}_a&\mathcal{A}_b\\\mathcal{B}_a&\mathcal{B}_b
\end{pmatrix}\gamma^G\begin{pmatrix}
\mathcal{A}_a&\mathcal{A}_b\\\mathcal{B}_a&\mathcal{B}_b
\end{pmatrix}^T\right]^{-1}= \left[\begin{pmatrix}
\mathcal{A}_a&\mathcal{A}_b\\\mathcal{B}_a&\mathcal{B}_b
\end{pmatrix}\mathcal M\gamma_{\mathrm{vac}}\mathcal M^T\begin{pmatrix}
\mathcal{A}_a&\mathcal{A}_b\\\mathcal{B}_a&\mathcal{B}_b
\end{pmatrix}^T\right]^{-1}.
\end{equation}
Thus,  Eq. (\ref{eigen}) expresses that  $\ket{\psi^G}$ is an eigenvector of ${1\over 2} \,{\bf R} ^T\mathbf{\Gamma}^{-1}{\bf R}$ with eigenvalue $n$, as advertized.

	\section*{References}
\bibliographystyle{unsrt}
\bibliography{references}

\end{document}